\pdfoutput=1
\documentclass[reprint,amsmath,amssymb,aps,prl,superscriptaddress,longbibliography,nobibnotes]{revtex4-2}
\usepackage{graphicx}%
\usepackage{dcolumn}%
\usepackage{bm}%

\usepackage{float}
\usepackage{framed}
\usepackage{physics}
\usepackage{amsfonts}
\usepackage[utf8]{inputenc}
\usepackage[T1]{fontenc}
\usepackage{amsmath}
\usepackage{textalpha}
\usepackage{newtxtext,newtxmath} %
\DeclareUnicodeCharacter{FF08}{(}

\usepackage{xcolor}

\newcommand{\LCN}{London Centre for Nanotechnology, University College London, 17-19 Gordon Street, London WC1H 0AH, UK}
\newcommand{\UCLEEE}{Department of Electronic and Electrical Engineering, University College London, 17-19 Gordon Street, London WC1H 0AH, UK}
\newcommand{\FRIS}{Frontier Research Institute for Interdisciplinary Sciences, Tohoku University, 6-3 Aoba, Sendai 980-8578, Japan}
\newcommand{\RIEC}{Research Institute of Electrical Communication, Tohoku University, 2-1-1 Katahira, Sendai 980-8577, Japan}
\newcommand{\NPL}{National Physical Laboratory, Teddington TW11 0LW, UK}
\newcommand{\CSIS}{Center for Science and Innovation in Spintronics, Tohoku University, 2-1-1 Katahira, Sendai 980-8577, Japan}
\newcommand{\IMR}{Institute for Materials Research, Tohoku University, 2-1-1 Katahira, Sendai 980-8577, Japan}
\newcommand{\AIMR}{WPI Advanced Institute for Materials Research, Tohoku University, 2-1-1 Katahira, Sendai 980-8577, Japan}

\begin{document}
\preprint{APS/123-QED}

\title{Giant magnetoresistance rectification by N\'eel vector dynamics}

\author{Zhiyi Che}
\thanks{These authors contributed equally to this work.}
\affiliation{\LCN}
\affiliation{\UCLEEE}

\author{Yifang Hao}
\thanks{These authors contributed equally to this work.}
\affiliation{\LCN}
\affiliation{\UCLEEE}

\author{Aakanksha Sud}
\affiliation{\FRIS}
\affiliation{\RIEC}

\author{Murat Cubukcu}
\affiliation{\LCN}
\affiliation{\NPL}

\author{Charlie W. F. Freeman}
\affiliation{\LCN}
\affiliation{\UCLEEE}
\affiliation{\NPL}

\author{Chenghao Yang}
\affiliation{\LCN}
\affiliation{\UCLEEE}

\author{Daniel Prestwood}
\affiliation{\LCN}
\affiliation{\UCLEEE}

\author{Troy Dion}
\affiliation{\CSIS}
\affiliation{\IMR}

\author{Shunsuke Fukami}
\affiliation{\RIEC}
\affiliation{\CSIS}
\affiliation{\AIMR}

\author{Shigemi Mizukami}
\affiliation{\CSIS}
\affiliation{\AIMR}

\author{Hidekazu Kurebayashi}
\email{h.kurebayashi@ucl.ac.uk}
\affiliation{\LCN}
\affiliation{\UCLEEE}
\affiliation{\CSIS}
\affiliation{\IMR}
\affiliation{\AIMR}

\date{\today}

\begin{abstract}
We observe room-temperature giant magnetoresistance (GMR) rectification driven by N\'eel vector dynamics in synthetic antiferromagnets. During spin-torque ferromagnetic resonance, the out-of-phase (optical) mode makes the relative angle between the two sub-moments oscillate, producing a rectified voltage due to GMR. This component is isolated from anisotropic and spin Hall magnetoresistance contributions by its angular symmetry, with its voltages scaling linearly with microwave power. We extract a GMR coefficient of 0.18--0.20\% from the rectification, which agrees with independently measured dc magnetoresistance. Reversing the N\'eel vector with a dc current changes the sign of the GMR rectification voltage in a non-volatile and controlled manner.
\end{abstract}

\maketitle
\textit{Introduction}---
Rectification, one of the oldest nonlinear effects in solids, was first reported by Braun in 1874, in point contacts on metal sulfides~\cite{Braun1874} later used as crystal radio detectors.
Its modern forms range from the semiconductor $p$--$n$ junction to optical rectification and second-harmonic generation, in which light mixes with itself at second order~\cite{sze_physics_2007,Franken1961,Bass1962,Boyd2020}.
Rikken et al.~\cite{Rikken2001} showed that rectification does not require a junction but can arise in the bulk, where the resistance of chiral bismuth helices under a magnetic field $\mathbf{B}$ gains a term proportional to $\mathbf{I}\cdot\mathbf{B}$, which changes sign with the current $\mathbf{I}$ and with the handedness of the helix.
This nonreciprocal transport has since been found in a wide range of materials~\cite{Tokura2018}, including magnetic bilayers~\cite{Avci2015} and superconductors, where it gives superconducting diodes~\cite{Ando2020,Nadeem2023}, while a Berry-curvature dipole produces a nonlinear Hall effect even without a magnetic field~\cite{Sodemann2015,Ma2019}.

Rectification also occurs in magnetic conductors, through magnetisation dynamics.
Current-induced torques from an rf current make the magnetisation precess, and magnetoresistance turns the precession into a resistance oscillation $\delta R(t)$ at the drive frequency.
Mixing $\delta R(t)$ with the current gives a dc voltage $V_{\mathrm{dc}} = \tfrac{1}{2}\,\delta R\, I_{\mathrm{rf}}\cos\varphi$, where $\varphi$ is the phase between current and resistance~\cite{Tulapurkar2005,Harder2016}.
Each step is linear in the current, so $V_{\mathrm{dc}}$ is second order, and it is a phase-sensitive readout of the magnetic motion with a magnitude set by $\delta R$.
This measurement technique, known as spin-torque ferromagnetic resonance (ST-FMR), has become a quantitative probe of current-induced torques, because the symmetric and antisymmetric Lorentzian components of the rectified voltages distinguish damping-like and field-like torques~\cite{Fang2011,Liu2011,Harder2016}. In these measurements, the resistance oscillation usually arises from the anisotropic magnetoresistance (AMR) or spin Hall magnetoresistance (SMR) of a single ferromagnetic layer~\cite{Liu2011,Liu2012,Fang2011,Kurebayashi2014,Mellnik2014,MacNeill2017,Chiba2014,Schreier2015,Sklenar2015}.
\begin{figure}[b]
    \centering
    \includegraphics[width=1.0\linewidth]{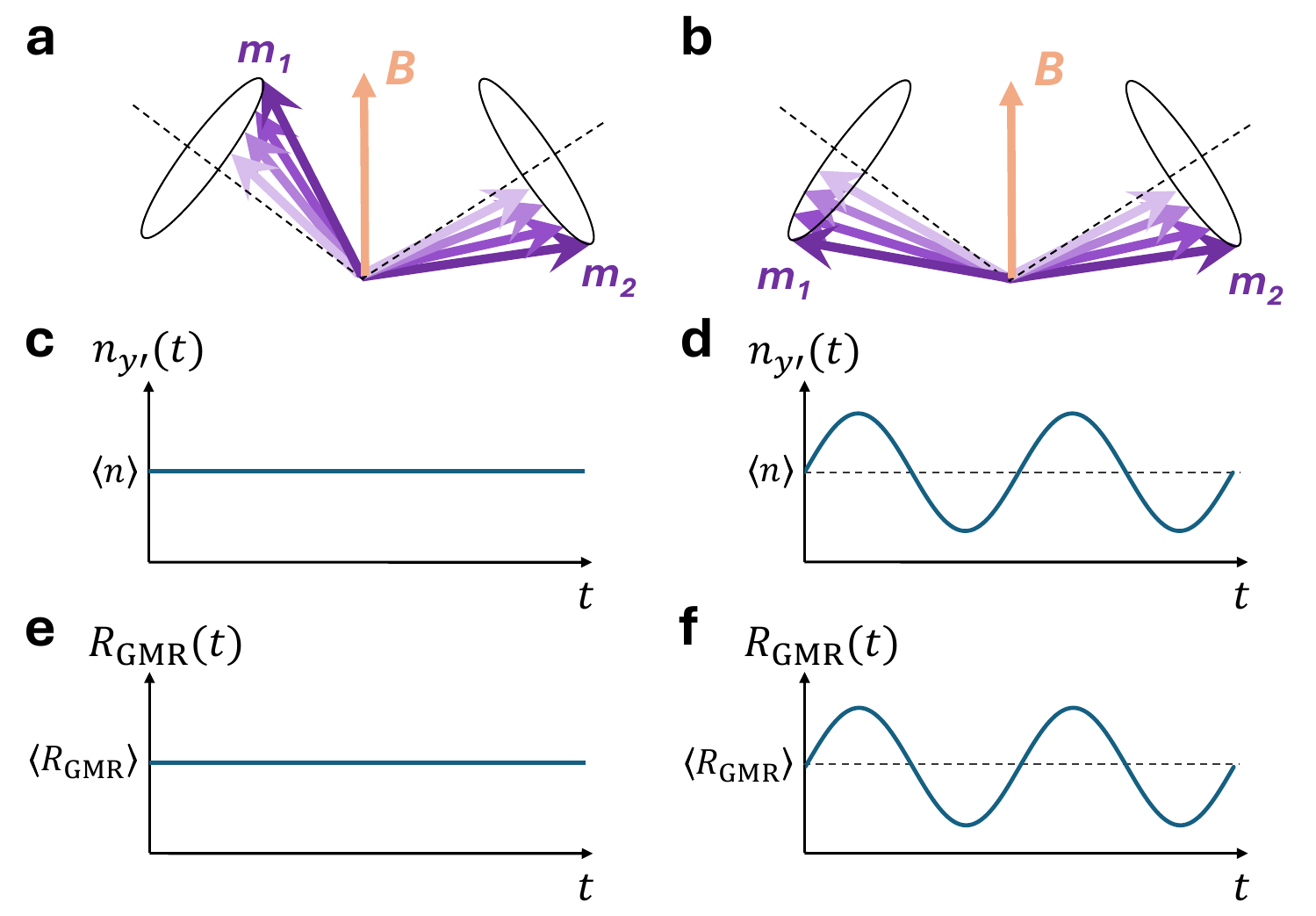}
    \caption{\textbf{Schematic of magnetisation dynamics and GMR rectification.} \textbf{a},\textbf{b}, Two coupled moments $\mathbf{m}_1$ and $\mathbf{m}_2$ precess in phase in the acoustic mode (\textbf{a}) and out of phase in the optical mode (\textbf{b}) under the applied magnetic field $\mathbf{B}$. \textbf{c},\textbf{d}, Time dependence of $n_{y'}=\mathbf{n}\cdot(\hat{\mathbf{z}}\times\mathbf{B})$, the component of the N\'eel vector perpendicular to the field, where $\langle n\rangle$ is the time-average of $\mathbf{n}$. To first order, $n_{y'}$ stays constant in the acoustic mode (\textbf{c}), but oscillates in the optical mode (\textbf{d}). \textbf{e},\textbf{f}, The resulting GMR resistance $R_\mathrm{GMR}(t)$ remains constant in the acoustic mode (\textbf{e}) but oscillates at the precession frequency in the optical mode (\textbf{f}).}
    \label{fig:Schematic}
\end{figure}

\begin{figure*}[t]
    \centering
    \includegraphics[width=1.0\linewidth]{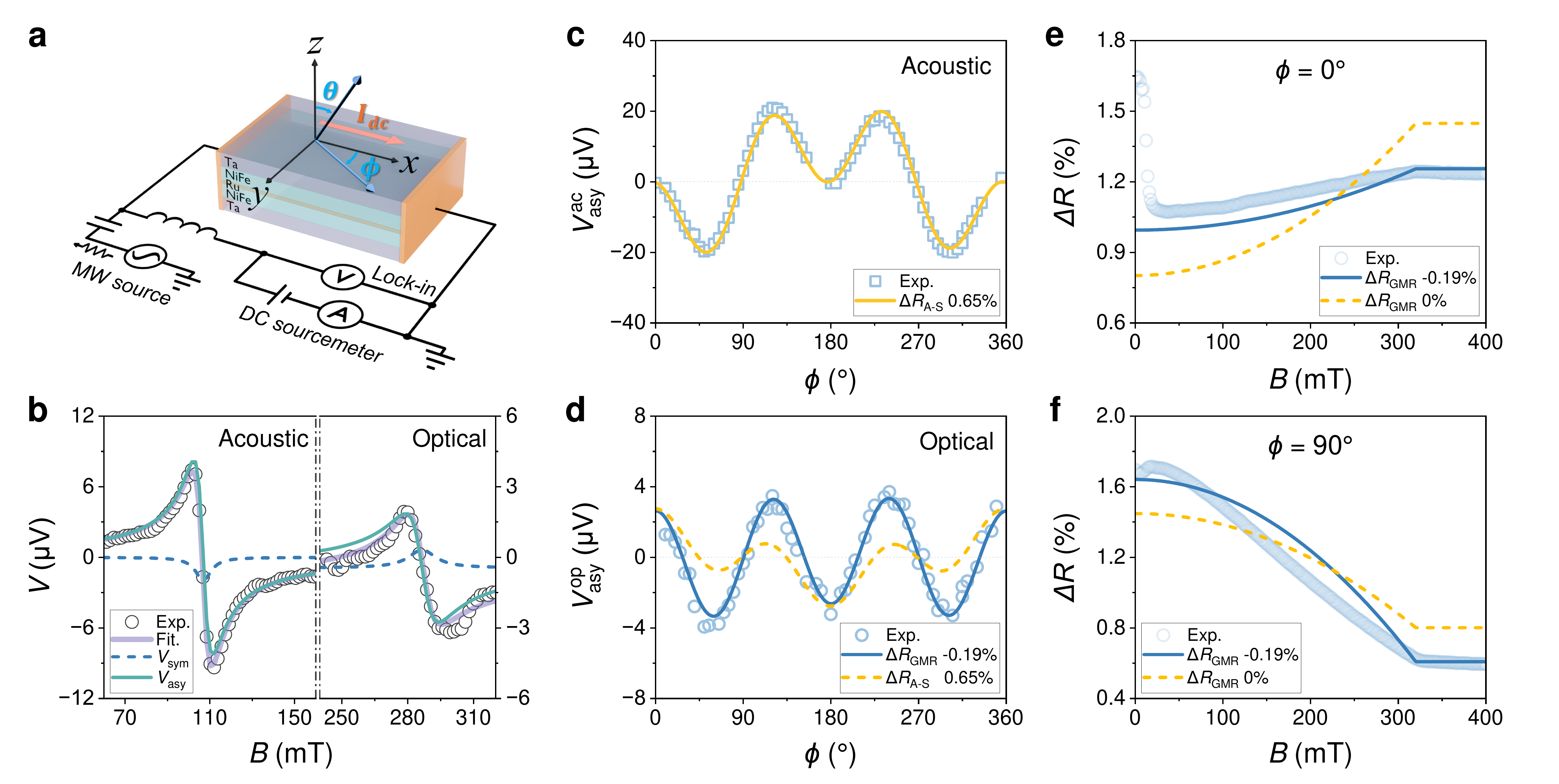}
    \caption{\textbf{Experimental setup and GMR rectification via in-plane magnetic resonance and resistance measurements.} 
    \textbf{a}, Schematic figure of experimental circuits and sample geometry. The sample is mounted on a coplanar waveguide (CPW) and measured by ST-FMR (microwave source and lock-in amplifier) or dc MR (sourcemeter). ST-FMR is measured at 7~GHz, and both the microwave and the switching dc currents flow along the easy axis ($x$). Unless otherwise noted, the field is applied in the film plane ($\theta=0^\circ$, defined in \textbf{a}) at an angle $\phi$ to the current. 
    \textbf{b}, Fitting of ST-FMR spectra using the Lorentzian function for acoustic and optical modes to separate the symmetric and antisymmetric components. 
    Antisymmetric component of FMR voltage $V_\mathrm{asy}$ as a function of magnetic field angle $\phi$ for acoustic (\textbf{c}) and optical (\textbf{d}) modes, fitted without (dashed) and with (solid) the GMR term. Measured MR and curves calculated for different $\Delta R_\mathrm{GMR}$ at \textbf{e}, $\phi=0^\circ$ and \textbf{f}, $\phi=90^\circ$.}
    \label{fig:setup_FMR}
\end{figure*}

Unlike AMR and SMR, giant magnetoresistance (GMR), discovered in antiferromagnetically coupled Fe/Cr multilayers~\cite{Baibich1988,Binasch1989}, requires two ferromagnetic moments and varies only with the angle between them.
Its rectification therefore depends not only on the phase between resistance and current but also on the relative phase of the two precessions: moments precessing in phase keep their relative angle fixed, to first order, and give no time-dependent GMR.
A synthetic antiferromagnet, two ferromagnetic layers coupled antiferromagnetically across a nonmagnetic spacer~\cite{duine2018synthetic}, offers a tunable model for antiferromagnetic spintronics~\cite{Jungwirth2016,Baltz2018}.
Its interlayer exchange field, first identified from its spin-wave modes~\cite{Grunberg1986}, splits the resonance into an acoustic branch (Fig.~\ref{fig:Schematic}a), in which the magnetisations precess in phase, and an optical branch (Fig.~\ref{fig:Schematic}b), in which they precess out of phase at a frequency raised by the coupling.
These branches have been measured by ST-FMR in synthetic antiferromagnets, with the rectified voltages analysed in terms of AMR and SMR~\cite{sud2020tunable,sud2021parity,chen2024electrical,sud2025electrically}.
Only the optical branch changes the angle between the two magnetisations, through the dynamics of the N\'eel vector $\mathbf{n}=(\mathbf{m}_1-\mathbf{m}_2)/2$, and so only it can produce GMR rectification.
Yet GMR has been used dynamically only to read the precession of a single free layer, in current-driven oscillators~\cite{Kiselev2003,Rossi2025,Li2025gmrshno} and by rectification~\cite{Sankey2006,Ziketek2015rectification}.

In this Letter, we report GMR rectification driven by N\'eel vector dynamics in a synthetic antiferromagnet.
In Ta(5)/NiFe(5)/Ru(0.35)/NiFe(5)/Ta(5) multilayers grown by magnetron sputtering on thermally oxidised Si substrates (thicknesses in nm; see Fig.~\ref{fig:setup_FMR}a), an rf current exerts torques of opposite sign on the two NiFe layers and excites both branches, which we detect by ST-FMR~\cite{sud2021parity,sud2025electrically}.
A GMR contribution appears only in the optical branch, which we separate from the AMR and SMR contributions by its angular dependence. 
We confirm that both contributions scale linearly with microwave power, as expected for second-order rectification.
The GMR coefficient, 0.18--0.20\% in four devices, agrees with dc magnetoresistance, and reversing the N\'eel vector with a dc current~\cite{hao2026synthetic} reverses the GMR voltage in a non-volatile manner.
These results establish GMR as an all-metallic, room-temperature electrical probe of N\'eel vector dynamics.

\textit{GMR rectification mechanism}---
For current along $x$, the longitudinal resistance of the synthetic antiferromagnet has three magnetoresistive contributions: AMR and SMR, which depend on the orientation of each magnetisation $\mathbf{m}_i$~\cite{fan2025giant,nakayama2013spin,althammer2013quantitative,chen2013theory}, and GMR, which depends only on their relative orientation; Hall effects do not contribute to the longitudinal voltage.
In the macrospin approximation~\cite{xiao2005macrospin},
\begin{multline}
\frac{\Delta R(t)}{R_0}=\frac{\Delta R_\mathrm{AMR}}{2}\sum_i m_{i,x}^2+\frac{\Delta R_\mathrm{SMR}}{2}\sum_i m_{i,y}^2\\
+\Delta R_\mathrm{GMR}\,\mathbf{m}_1\cdot\mathbf{m}_2,
\label{eq:R}
\end{multline}
where $\Delta R_\mathrm{AMR}$, $\Delta R_\mathrm{SMR}$ and $\Delta R_\mathrm{GMR}$ are the AMR, SMR and GMR resistance changes, respectively. By contrast, the ordinary GMR resistance change defined by the parallel and antiparallel states differs by $2|\Delta R_\mathrm{GMR}|$.
Because AMR and SMR also change as the layers cant, the field dependence of $R$ alone does not isolate GMR.
We therefore measure $\Delta R_\mathrm{AMR}$ and $\Delta R_\mathrm{SMR}$ independently by rotating a saturating field out of plane~\cite{SM}, leaving the GMR coefficient $\Delta R_\mathrm{GMR}$ as the only free magnetoresistance parameter.

Near each resonance, the rectified voltage takes the Lorentzian form~\cite{Liu2011,sud2020tunable,sud2021parity}
\begin{multline}
V_\mathrm{dc}=V_0+V_\mathrm{sym}\frac{\Delta B^2}{(B-B_\mathrm{res})^2+\Delta B^2}\\
+V_\mathrm{asy}\frac{\Delta B\,(B-B_\mathrm{res})}{(B-B_\mathrm{res})^2+\Delta B^2},
\label{eq:Lorentz}
\end{multline}
where $V_0$ is a background voltage, $B_\mathrm{res}$ the resonance field, $\Delta B$ the half width at half maximum, and $V_\mathrm{sym}$ and $V_\mathrm{asy}$ the amplitudes of the symmetric and antisymmetric components. Figure~\ref{fig:setup_FMR}b shows fits of the ST-FMR spectra with Eq.~\eqref{eq:Lorentz}.
Linearising Eq.~\eqref{eq:R} about the canted equilibrium with the coupled macrospin equations of motion~\cite{zhang1994angular,SM} gives the antisymmetric amplitudes
\begin{align}
V^\mathrm{ac}_\mathrm{asy}&=-\frac{I_\mathrm{rf}R_0B_\mathrm{FL}\,\langle n\rangle(1-2\langle n\rangle^2)}{8\Delta B_\mathrm{ac}\sqrt{1-\langle n\rangle^2}}\nonumber\\
&\quad\times(\Delta R_\mathrm{AMR}-\Delta R_\mathrm{SMR})\sin2\phi\sin\phi, \label{eq:Vasy_ac}\\
V^\mathrm{op}_\mathrm{asy}&=-\frac{I_\mathrm{rf}R_0B_\mathrm{FL}\,\langle n\rangle\sqrt{1-\langle n\rangle^2}}{4\Delta B_\mathrm{op}}\nonumber\\
&\quad\times\big[(\Delta R_\mathrm{AMR}-\Delta R_\mathrm{SMR})\cos2\phi+2\Delta R_\mathrm{GMR}\big]\cos\phi, \label{eq:Vasy_op}
\end{align}
where $I_\mathrm{rf}$ is the rf current amplitude, $B_\mathrm{FL}$ the effective field of the field-like torque, $\phi$ the angle between field and current, $\langle n\rangle$ the component of the equilibrium N\'eel vector along $\hat{\mathbf{z}}\times\hat{\mathbf{B}}$, whose sign distinguishes the two canted states, $\Delta B_\mathrm{ac(op)}$ the linewidth of each branch, and the superscripts ac and op denote the acoustic and optical branches.
The symmetric amplitudes, set by the damping-like torque, have the same angular dependences~\cite{SM}.
GMR enters only the optical branch because in-phase precession leaves $\mathbf{m}_1\cdot\mathbf{m}_2$ unchanged to first order (Fig.~\ref{fig:Schematic}c,e), whereas out-of-phase precession modulates it through the N\'eel vector (Fig.~\ref{fig:Schematic}d,f).
The GMR term also has a distinct angular dependence, $\cos\phi$ rather than $\cos2\phi\cos\phi$, so the two contributions can be separated; at $\phi=45^\circ$, GMR is the only one.
Its prefactor $\langle n\rangle\sqrt{1-\langle n\rangle^2}$ vanishes in both the collinear and the saturated state, so GMR rectification requires the canted state.

\textit{Experimental signature}---
The GMR rectification in the optical mode is evidenced by the angular dependence of $V_\mathrm{asy}$. 
We measure ST-FMR at 7\,GHz and room temperature, with the field applied in the film plane at angle $\phi$ to the current~\cite{SM}, and analyse $V_\mathrm{asy}$. 
The acoustic mode follows $\sin2\phi\sin\phi$ (Fig.~\ref{fig:setup_FMR}c), as expected from AMR and SMR alone, which enter Eq.~\eqref{eq:Vasy_ac} only through $\Delta R_\mathrm{A-S}=\Delta R_\mathrm{AMR}-\Delta R_\mathrm{SMR}$; no $\Delta R_\mathrm{GMR}$ term is needed.
We fix $\Delta R_\mathrm{A-S}=0.65\%$ from the in-plane angular dependence of MR under a saturating field (Fig.~\ref{fig:SI_OOP_MR}a), in agreement with the difference between $\Delta R_\mathrm{AMR}=1.45\%$ and $\Delta R_\mathrm{SMR}=0.8\%$ at $B=3\,\mathrm{T}$ measured by rotating a 3\,T field out of plane (Fig.~\ref{fig:SI_OOP_MR}b-d).  
However, the $\cos2\phi\cos\phi$ curve expected from AMR and SMR visibly fails to reproduce $V_\mathrm{asy}^\mathrm{op}$ (dashed line, Fig.~\ref{fig:setup_FMR}d), pointing to an additional mechanism.
Because $\Delta R_\mathrm{A-S}$ is fixed, $\Delta R_\mathrm{GMR}$ is set by the shape of the angular dependence rather than its amplitude, and is therefore independent of $B_\mathrm{FL}$, the linewidth and the rf-current calibration.
The symmetric components likewise require the GMR term in the optical mode (Fig.~\ref{fig:SI_FMR_sym}).
Repeating this analysis for four devices, two each with $t_\mathrm{Ru}=0.35$ and 0.40\,nm, demonstrates the reproducibility: $\Delta R_\mathrm{GMR}$ ranges from $-0.20\%$ to $-0.18\%$, while $\Delta R_\mathrm{A-S}$ ranges from 0.60\% to 0.95\% (Fig.~\ref{fig:SI_device_vs}).

The GMR contribution is confirmed independently by the dc magnetoresistance.  
As shown in Fig.~\ref{fig:setup_FMR}e,f, the MR curve calculated without GMR deviates from the data for both field-sweep directions (see SM~\cite{SM} for further characterisation), showing that AMR and SMR alone cannot account for the measured magnetoresistance. Fitting the MR curves with $\Delta R_\mathrm{GMR}$ as the only free parameter gives $\Delta R_\mathrm{GMR}=-0.19$\%, close to the value from rectification, and the same agreement holds at other $\phi$ (Fig.~\ref{fig:SI_MR_calc}).
The resistance jumps near zero field can be attributed to domain formation and propagation during magnetisation reversal~\cite{yan2023topological,zhao2023ferromagnetism}, which our macrospin model does not capture. 
The small differences between the calculated and measured curves may arise from the shape anisotropy, which our model neglects~\cite{camley1989theory}.

\begin{figure}[t]
    \centering
    \includegraphics[width=1.0\linewidth]{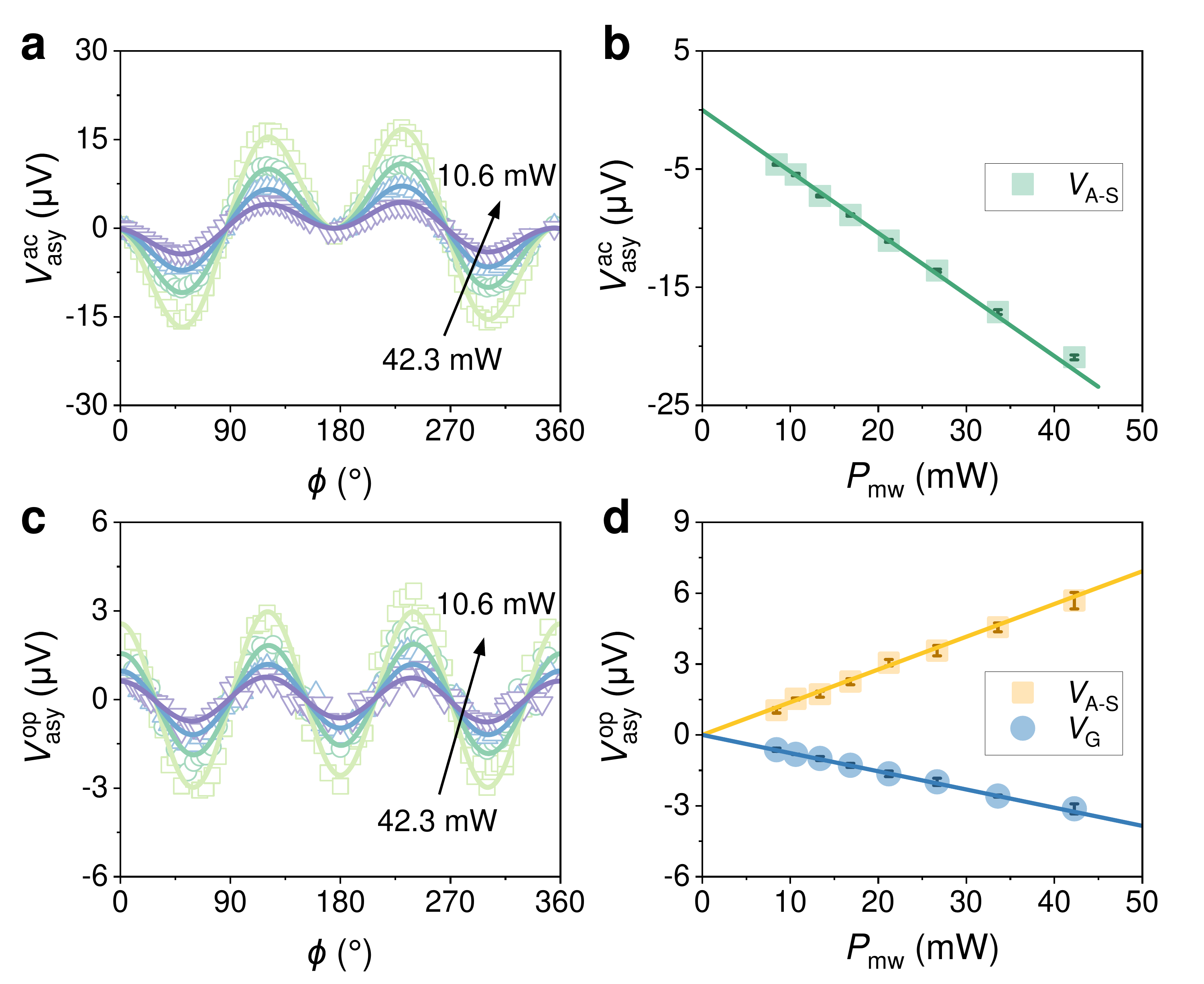}
    \caption{\textbf{Power dependence of MR effects clarified by magnetic resonance.} $V_\mathrm{asy}$ as a function of field angle $\phi$ at several microwave powers for the acoustic (\textbf{a}) and optical (\textbf{c}) modes. $V_\mathrm{A-S}$ and $V_\mathrm{G}$ as a function of microwave power $P_\mathrm{mw}$ for the acoustic (\textbf{b}) and optical (\textbf{d}) modes.}
    \label{fig:mw_asy}
\end{figure}
\textit{Quantification of GMR rectification}---
The GMR rectification is further quantified from its microwave power dependence.
We calibrate the rf current $I_\mathrm{rf}$ in the device bolometrically, by comparing the resistance change from Joule heating by a dc current with that by the microwave current (Fig.~\ref{fig:SI_mw_calibration}), which gives the power delivered to the device, $P_\mathrm{mw}$.
Since $B_\mathrm{FL}\propto I_\mathrm{rf}$~\cite{SM}, Eqs.~\eqref{eq:Vasy_ac} and \eqref{eq:Vasy_op} predict $V_\mathrm{asy}\propto I_\mathrm{rf}^2\propto P_\mathrm{mw}$, as expected for a second-order response.
Since the GMR readout varies as $\cos\phi$, whereas the combined AMR and SMR readout varies as $\sin2\phi\sin\phi$ in the acoustic mode and $\cos2\phi\cos\phi$ in the optical mode, $V_\mathrm{A-S}$ and $V_\mathrm{G}$ can be separated from $V_\mathrm{asy}$ (Supplementary Note~2~\cite{SM}). Their ratio, $V_\mathrm{G}/V^\mathrm{op}_\mathrm{A-S}=2\Delta R_\mathrm{GMR}/\Delta R_\mathrm{A-S}$, should therefore be independent of power.

The angular dependence of the FMR voltage is measured for both the acoustic and optical modes at $P_\mathrm{mw}$ from 8.4 to 42.3\,mW (Fig.~\ref{fig:mw_asy}a,c). 
We use the full angular dependence, rather than the voltages at two angles, because each fit parameter is then determined from many data points, which reduces the effect of measurement noise when quantifying $V_\mathrm{G}$ and $V^\mathrm{op}_\mathrm{A-S}$.
In the acoustic mode, $\sin2\phi\sin\phi$ describes the data at every power without a GMR term, and $V^\mathrm{ac}_\mathrm{A-S}$ increases linearly in magnitude with $P_\mathrm{mw}$ (Fig.~\ref{fig:mw_asy}b).
In the optical mode, $V^\mathrm{op}_\mathrm{A-S}$ and $V_\mathrm{G}$ both increase linearly in magnitude with $P_\mathrm{mw}$, with opposite signs (Fig.~\ref{fig:mw_asy}d).
Their ratio is constant, $V_\mathrm{G}/V^\mathrm{op}_\mathrm{A-S}=-0.56$, giving $\Delta R_\mathrm{GMR}=-0.18\%$, consistent with the single-power analysis in Fig.~\ref{fig:setup_FMR}d.
The GMR voltage therefore scales with power in the same way as the AMR and SMR voltages of the same mode, as expected if both arise from the same optical-mode precession.
The symmetric components show the same linear dependence (Fig.~\ref{fig:SI_mw_sym}).

\begin{figure}[t]
    \centering
    \includegraphics[width=1.0\linewidth]{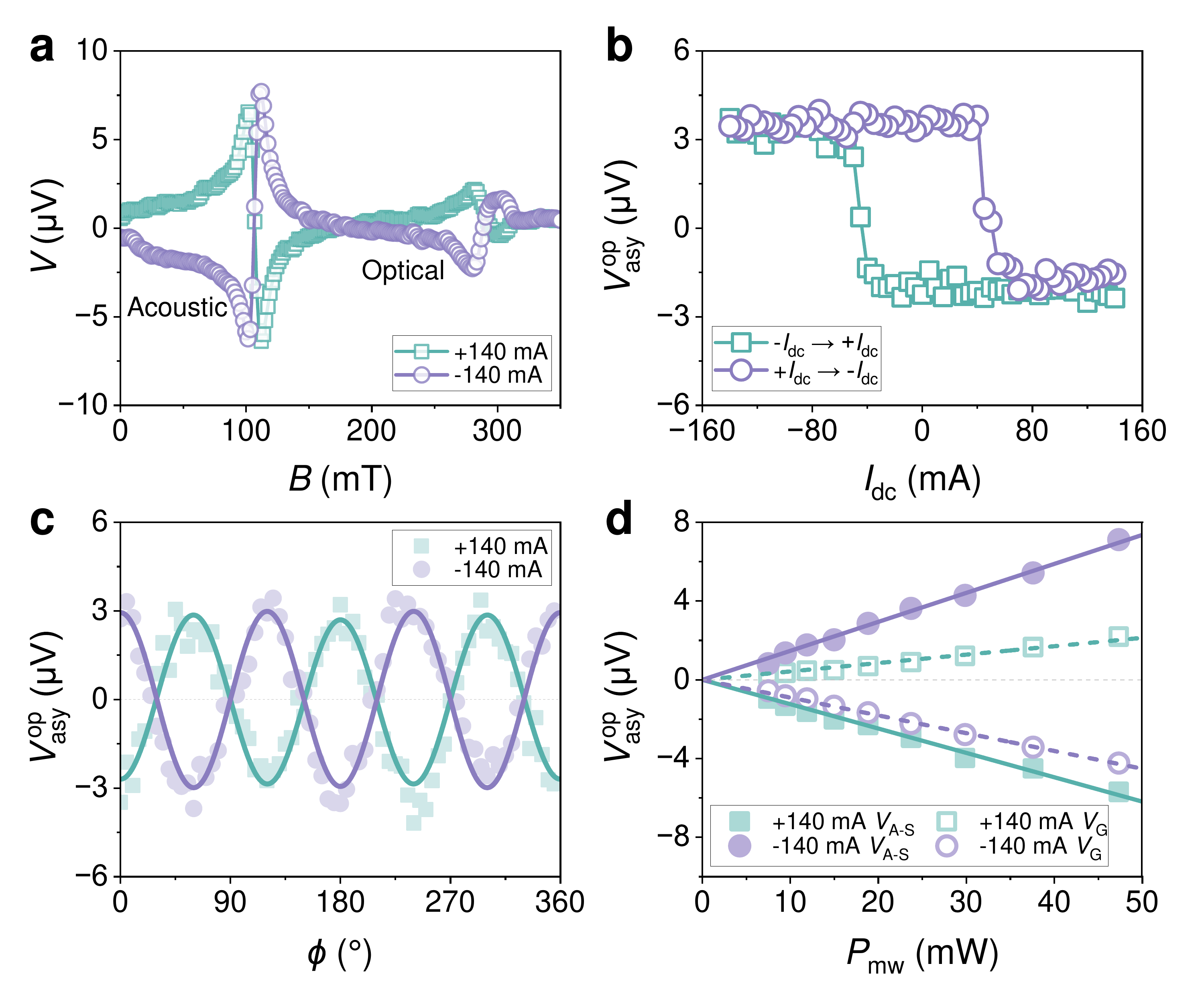}
    \caption{\textbf{GMR rectification of current-induced N\'eel vector reversal.} \textbf{a}, FMR voltage as a function of magnetic field after positive (green) and negative (purple) switching currents $I_\mathrm{dc}$. \textbf{b}, $V^\mathrm{op}_\mathrm{asy}$ during $I_\mathrm{dc}$ sweeps. \textbf{c}, $V^\mathrm{op}_\mathrm{asy}$ as a function of field angle $\phi$ for both states. \textbf{d}, $V^\mathrm{op}_\mathrm{A-S}$ (solid symbols) and $V_\mathrm{G}$ (open symbols) as a function of microwave power $P_\mathrm{mw}$ for both states.}
    \label{fig:dc_mw_asy}
\end{figure}
\textit{Electrical control of GMR rectification}--- In our previous work on the same synthetic antiferromagnet, a dc current reversed the N\'eel vector~\cite{hao2026synthetic}. The current exerts torques of opposite sign on the two NiFe layers, which set the sign of $\mathbf{n}$, and the state is retained after the current is removed. The rf current in ST-FMR also exerts torques of opposite sign on the two layers, so the rectified voltages change sign when $\mathbf{n}$ is reversed. The GMR voltage should therefore reverse with the N\'eel vector, while $\Delta R_\mathrm{GMR}$ remains unchanged.

To demonstrate electrical control of GMR rectification, we apply a positive or negative dc current $I_\mathrm{dc}$ to the device and remove it 10~s before each FMR measurement. Figure~\ref{fig:dc_mw_asy}a shows field sweeps of the FMR voltage after $I_\mathrm{dc}=\pm140$~mA, with clear sign changes in both the acoustic and optical resonances. Measuring the $I_\mathrm{dc}$ dependence at the optical resonance in more detail, we find that $V^\mathrm{op}_\mathrm{asy}$ extracted from each trace forms a hysteresis loop (Fig.~\ref{fig:dc_mw_asy}b), indicating non-volatile electrical reversal of the N\'eel vector. To confirm that the electrically set N\'eel states are non-volatile and robust against magnetic fields, we apply $I_\mathrm{dc}$ only once, at $\phi=20^\circ$, and then measure the full angular dependence (Fig.~\ref{fig:dc_mw_asy}c), sweeping the field back and forth through the optical resonance at each angle.

We obtain $V^\mathrm{op}_\mathrm{A-S}$ and $V_\mathrm{G}$ in each state from the antisymmetric amplitudes at $\phi=120^\circ$ and $140^\circ$.
Equation~\eqref{eq:Vasy_op} has the form $V^\mathrm{op}_\mathrm{asy}(\phi)=(V^\mathrm{op}_\mathrm{A-S}\cos2\phi+V_\mathrm{G})\cos\phi$, which gives
\begin{align}
V^\mathrm{op}_\mathrm{A-S}&\approx2.97\,V^\mathrm{op}_\mathrm{asy}(120^\circ)-1.94\,V^\mathrm{op}_\mathrm{asy}(140^\circ),\nonumber\\
V_\mathrm{G}&\approx-0.52\,V^\mathrm{op}_\mathrm{asy}(120^\circ)-0.97\,V^\mathrm{op}_\mathrm{asy}(140^\circ).\nonumber
\end{align}
We extract both components from field sweeps at each microwave power $P_\mathrm{mw}$ delivered to the device (see SM for the calibration), and Fig.~\ref{fig:dc_mw_asy}d shows the results for $I_\mathrm{dc}=\pm140$~mA.
Both $V^\mathrm{op}_\mathrm{A-S}$ and $V_\mathrm{G}$ increase linearly with $P_\mathrm{mw}$ and change sign when $I_\mathrm{dc}$ is reversed. 
The sign of the GMR rectification voltage is therefore set by the static N\'eel vector, which can be reversed electrically.

For these measurements, we notice that there exists an asymmetry in amplitude (also noticeable in Fig.~\ref{fig:dc_mw_asy}b) before and after the N\'eel vector reversal.
If the two NiFe layers and the torques acting on them were identical, the two N\'eel states set by dc currents of opposite polarity would produce $V^\mathrm{op}_\mathrm{A-S}$ and $V_\mathrm{G}$ of equal magnitude. 
Any difference between the two layers breaks the inversion symmetry of the stack and can cause an amplitude difference after the N\'eel vector reversal.
Differences in saturation magnetisation, magnetoresistance or current-induced torques are possible sources, and such differences are expected because the two NiFe layers are grown on different underlayers (Ta and Ru).
We emphasise, however, that this asymmetry only adds complexity to the model analysis and does not affect our central claim of GMR rectification by N\'eel vector dynamics and its electrical control, since both $V^\mathrm{op}_\mathrm{A-S}$ and $V_\mathrm{G}$ reverse sign with the N\'eel vector. 

\textit{Conclusion}---
In summary, we observe GMR rectification driven by N\'eel vector dynamics in a NiFe/Ru/NiFe synthetic antiferromagnet at room temperature.
Because GMR depends only on the relative orientation of the two magnetisations, the GMR voltage appears in the optical mode and is absent in the acoustic mode.
Its $\cos\phi$ angular dependence separates it from the AMR and SMR voltages and gives a GMR coefficient of 0.18--0.20\% in four devices, close to the 0.19\% obtained from dc magnetoresistance.
The GMR voltage scales linearly with microwave power, as expected for second-order rectification.
A dc current pulse reverses the N\'eel vector that causes a sign change of the GMR rectification voltage in a non-volatile manner.
GMR rectification therefore reads both the dynamics and the sign of the N\'eel vector, meeting a central requirement of antiferromagnetic spintronics~\cite{Jungwirth2016,Baltz2018}.
Because out-of-phase modes are fundamental to any coupled-moment system, the presented mechanism should respond to them in antiferromagnets, albeit at high frequencies set by the exchange interaction in crystalline antiferromagnets~\cite{Baltz2018}. 
Layered van der Waals antiferromagnets with modest interlayer coupling are more accessible due to their GHz resonance frequencies, such as CrSBr~\cite{Bae_Nature2022,Cham_Science2025}. A similar N\'eel-vector-driven rectification mechanism should exist in a tunnelling geometry using insulating CrI$_3$ and CrCl$_3$~\cite{Song_Science2018,macneill2019gigahertz}.

\textit{Acknowledgments}—
H. K. acknowledges Engineering and Physical Sciences
Research Council for support via grant EP/X015661/1, and
UCL for UCL-Tohoku University Strategic Partner Fund.
S. F. thanks to JSPS KAKENHI (No. 24H00039,
No. 24H02235). A. S. thanks JSPS Postdoctoral
fellowship for research in Japan (P21777). 
S. M. thanks CSRN in CSIS at Tohoku University and JSPS
KAKENHI (No. 21H04648, No. 21H05000,
No. 22F21777, No. 22KF0030, No. 26H02126) and X-NICS of MEXT
(No. JPJ011438). M.C. acknowledges support from the UK
Government Department for Science, Innovation and Technology through
National Measurement System (NMS) funding (semiconductor metrology).

\textit{Data availability}—
The data that support the findings of
this article are not publicly available. The data are available
from the authors upon reasonable request.
\bibliography{references}%

\clearpage
\onecolumngrid
\setcounter{equation}{0}
\setcounter{figure}{0}
\setcounter{table}{0}
\renewcommand{\theequation}{S\arabic{equation}}
\renewcommand{\thefigure}{S\arabic{figure}}
\renewcommand{\thetable}{S\arabic{table}}
\begin{center}
{\large\bfseries Supplemental Material for ``Giant magnetoresistance rectification by N\'eel vector dynamics''}\\[8pt]
Zhiyi Che, Yifang Hao, Aakanksha Sud, Murat Cubukcu, Charlie W. F. Freeman, Chenghao Yang, Daniel Prestwood, Troy Dion, Shunsuke Fukami, Shigemi Mizukami, and Hidekazu Kurebayashi
\end{center}
\vspace{12pt}
\begin{center} {\Large CONTENTS} \end{center}
{\large Supplementary Note 1: Phenomenological models}\\
{\large Supplementary Note 2: Power dependence of magnetoresistance effects}\\
{\large Supplementary Note 3: GMR rectification of current-induced N\'eel vector reversal}\\
{\large Supplementary Figures S4-S9}\\
\newpage

\section*{Supplementary Note 1: Phenomenological models}

\subsection{Magnetic free energy for the canted state}
Magnetic free energy of the synthetic antiferromagnetic (SAF) system can be derived by the Zeeman, demagnetisation, and interlayer-exchange terms: 
\begin{align}
    E = \sum_{j=1}^{2}\left[ M_\mathrm{S} B\cdot m_j+\frac{1}{2}M_\mathrm{S} B_\mathrm{S}(m_j\cdot z)^2 \right] + \frac{2J_\mathrm{ex,1}}{d}m_1\cdot m_2 + \frac{2J_\mathrm{ex,2}}{d}(m_1\cdot m_2)^2
    \label{eq:SI_free_energy}
\end{align}
where $M_\mathrm{S}$, $m_i$, $B$, $B_\mathrm{S}$, $J_\mathrm{ex,i}$, $d$ are the the saturation magnetisation of the system, the unit vector of moment in each ferromagnetic layer, external magnetic field vector, demagnetisation field, the linear and quadratic antiferromagnetic interlayer exchange coupling constants, and the thickness of single ferromagnetic layer, respectively. Herein, the linear exchange field $B_\mathrm{ex}$ is 160\,mT, and the $M_\mathrm{S}=680\,\mathrm{kA/m}$ for Ru thickness of 0.35\,nm in this case, and $B_\mathrm{ex,2}$ is ignored to simplify the calculation due to $-J_\mathrm{ex,2}\ll J_\mathrm{ex,1}$. Hence, the linear exchange constant can be obtained from the exchange field $B_\mathrm{ex}=-J_\mathrm{ex,1}/(M_\mathrm{S}d)$.\\
The canted angle $\phi_\mathrm{c,i}$ between the equilibrium direction of individual moment $m_i$ and applied magnetic field $B$ is used in the macrospin model to describe the magnetic moment behaviour during external field sweep, and the demagnetisation term $m_j\cdot z$ is zero for in-plane field rotation. Therefore, the free energy expression can be reduced to:
\begin{align}
    \frac{E(\phi_\mathrm{c,1},\phi_\mathrm{c,2})}{M_\mathrm{S}} = -B(\cos\phi_\mathrm{c,1}+\cos\phi_\mathrm{c,2}) + B_\mathrm{ex}\cos(\phi_\mathrm{c,1}-\phi_\mathrm{c,2})
    \label{eq:SI_macrospin}
\end{align}
To minimise the energy, $\frac{\partial E}{\partial \phi_\mathrm{c,i}}=0$ is taken, and symmetric canted angle $\phi_\mathrm{c,1}=-\phi_\mathrm{c,2}=\phi_\mathrm{c}$ is inferred in the macrospin model. Solving the above functions, the expression $\cos\phi_\mathrm{c}(B)=\frac{B}{2B_\mathrm{ex}}$ is derived to illustrate the relation between canted angle $\phi_\mathrm{c}$ and applied field $B$ in canted state.\\

\subsection{FMR voltage rectifications in synthetic antiferromagnets}
The rectification voltages generated from magnetisation dynamics in synthetic antiferromagnets are established as following, where an alternative coordination with the field direction as $x'$axis ($x'=x\cos\phi+y\sin\phi$, $y'=-x\sin\phi+y\cos\phi$ and $z'=z$) is used.
For the magnetisations of two ferromagnetic sublattice $m_1=m^0_1+\delta m_1(t)$ and $m_2=m^0_2+\delta m_2(t)$, the Kittel and N\'eel vectors are defined as $m = \langle m\rangle+\delta m(t)=(m_1+m_2)/2$ and $n=\langle n\rangle+\delta n(t)=(m_1-m_2)/2$, respectively. The Landau-Lifshitz-Gilbert(LLG) equations adopting these coupled moments are rewritten as:
\begin{equation}
    \begin{aligned}
    \frac{dm}{dt} &= -\Omega_\mathrm{L} m\times u + \Omega_\mathrm{B}\Big[m\times\Big(m\cdot \hat{e}_z\Big)\hat{e}_z + n\times\Big(n\cdot \hat{e}_z\Big)\hat{e}_z\Big] + \alpha\left(m\times \frac{dm}{dt} + n\times \frac{dn}{dt}\right)
    \label{eq:SI_LLG_m}
    \end{aligned}
\end{equation}
\begin{equation}
\begin{aligned}
    \frac{dn}{dt} &= -\Omega_\mathrm{L} n\times u + \Omega_\mathrm{B}\Big[n\times\Big(m\cdot \hat{e}_z\Big)\hat{e}_z + m\times\Big(n\cdot \hat{e}_z\Big)\hat{e}_z\Big] + 2\Omega_\mathrm{ex}\Big(n\times m\Big)+\alpha\left(n\times \frac{dm}{dt} + m\times \frac{dn}{dt}\right)
    \label{eq:SI_LLG_n}
\end{aligned}
\end{equation}
where $\Omega_\mathrm{L}=\gamma B$, $\Omega_\mathrm{B}=\gamma B_\mathrm{s}$, $\Omega_\mathrm{ex}=\gamma B_\mathrm{ex}$ and $u$ are the unit vector along the magnetic field $B$ direction with the expression of $u=\hat{e}_{x'}$. 
\\
Two configurations of flipped N\'eel vectors are taken into account: $m^0_1=(\sqrt{1-\langle n\rangle^2},\; \langle n\rangle,\; 0),\; m^0_2=(\sqrt{1-\langle n\rangle^2},\; -\langle n\rangle,\; 0)$ with linearised $m$ and $n$ as $m=(
\sqrt{1-\langle n\rangle^2} + \delta m_{x'}(t),\; \delta m_{y'}(t),\;  \delta m_{z'}(t))$ and $n=(\delta n_{x'}(t),\; \langle n\rangle + \delta n_{y'}(t),\; \delta n_{z'}(t))$. Solving Eqs.~\eqref{eq:SI_LLG_m}-\eqref{eq:SI_LLG_n} for the first order term, the matrix of two equations is attained from Fourier transformation as following:
\begin{align}
\begin{pmatrix}
     i\Omega & (i\alpha\Omega + \Omega_\mathrm{B} )\sqrt{1-\langle n\rangle^2} + \Omega_\mathrm{L} & 0 \\
     -i\alpha\Omega \sqrt{1-\langle n\rangle^2} - \Omega_\mathrm{L} & i\Omega & i\alpha\Omega \langle n\rangle \\
     0 & -(i\alpha\Omega +2\Omega_\mathrm{ex}+\Omega_\mathrm{B})\langle n\rangle\ & i\Omega
\end{pmatrix}
\begin{pmatrix}
    \delta m_{y'}(t) \\ \delta m_{z'}(t) \\ \delta n_{x'}(t)
\end{pmatrix}
    = \bm{\tau}_m
    \label{eq:SI_matrix_m}
\end{align}
\begin{align}
\begin{aligned}
\begin{pmatrix}
     i\Omega & 0 & -(i\alpha\Omega + \Omega_\mathrm{B}) \langle n\rangle \\
     0 & i\Omega & (i\alpha\Omega - 2\Omega_\mathrm{ex}+\Omega_\mathrm{B})\sqrt{1-\langle n\rangle^2} + \Omega_\mathrm{L} \\
     (i\alpha\Omega + 2\Omega_\mathrm{ex})\langle n\rangle & (-i\alpha\Omega+2\Omega_\mathrm{ex}) \sqrt{1-\langle n\rangle^2} - \Omega_\mathrm{L} & i\Omega
\end{pmatrix}
\\[6pt]
\begin{pmatrix}
    \delta m_{x'}(t) \\ \delta n_{y'}(t) \\ \delta n_{z'}(t)
\end{pmatrix}
    = \bm{\tau}_n
\end{aligned}
\label{eq:SI_matrix_n}
\end{align}
The right-hand side terms correspond to magnetic torques acting on each coupled moment. The resonance frequencies of both modes can be extracted from the determinant of the $3\times3$ matrix on the left-hand side, and the relation $\sqrt{1-\langle n\rangle^2}=\Omega_\mathrm{L}/2\Omega_\mathrm{ex}$ is used for simplicity:
\begin{equation}
\begin{aligned}
    \tilde{\Omega}_\mathrm{ac} &= \sqrt{\Omega_\mathrm{L}^2\left(1+\frac{\Omega_\mathrm{B}}{2\Omega_\mathrm{ex}}\right)} + i\frac{\alpha}{2}\left(\frac{\Omega_\mathrm{L}^2}{2\Omega_\mathrm{ex}}+\Omega_\mathrm{B}+2\Omega_\mathrm{ex}\right)\\
    \tilde{\Omega}_\mathrm{op} &= \sqrt{\frac{\Omega_B}{2\Omega_\mathrm{ex}}(4\Omega_\mathrm{ex}^2 - \Omega_L^2)} + i\frac{\alpha}{2}\left(\Omega_B+\frac{1}{2\Omega_\mathrm{ex}}(4\Omega_\mathrm{ex}^2-\Omega_L^2)\right) 
\end{aligned}
\end{equation}
where the electrical current $I_\mathrm{rf}$ is applied along $x$ axis and magnetic field $B$ along $x'$.\\
In this alternative coordinate, the generated field-like effective field $B_\text{FL}$ exists in $y$ direction, and exerting field-like torques ($\tau_\mathrm{FL,1(2)}$) on $m_1$ and $m_2$ have the expression of:
\begin{align}
    \tau_\mathrm{FL,1(2)} &= \gamma B_{\mathrm{FL},1(2)} m_{1(2)} \times \hat{e}_{y} \nonumber\\
    &= \gamma B_{\mathrm{FL},1(2)} (\sqrt{1-\langle n\rangle^2}\cos\phi-\langle n\rangle\sin\phi)\, \hat{e}_{z'}
\end{align}
Especially, $B_{\mathrm{FL},1}=-B_{\mathrm{FL},2}$ for symmetric stacking. Therefore, the field-like torques acting on $m$ and $n$ in Eqs.~\eqref{eq:SI_matrix_m}–\eqref{eq:SI_matrix_n} are rewritten in:
\begin{align}
    \tau_m &= \gamma
    \begin{pmatrix}
        0 \\
        -B_\mathrm{FL}\,\langle n\rangle \sin\phi \\
        0
    \end{pmatrix}
    \label{eq:SI_tau_m_FL}
\end{align}
\begin{align}
    \tau_n &= \gamma
\begin{pmatrix}
    0 \\
    0 \\
    B_\mathrm{FL}\,\sqrt{1-\langle n\rangle^2} \cos\phi
    \label{eq:SI_tau_n_FL}
\end{pmatrix}
\end{align}
Besides, the damping-like torques generated from spin Hall effect (SHE) takes the form of $\tau_\mathrm{DL,i}= \gamma B_\mathrm{DL,i} \big[m_i \times (\hat{e}_{y} \times m_i)\big], (i=1,2)$ with the spin polarization $\sigma=\hat{e}_{y}$. The magnitude of $B_\mathrm{DL,i}$ is given by $B_\mathrm{DL}=\hbar\theta_\mathrm{SH}I_\mathrm{Ta}/2eM_\mathrm{S}wd_\mathrm{FM}d_\mathrm{Ta}$ where $\hbar$, $\theta_\mathrm{SH}$, $I_\mathrm{Ta}$ $e$, $M_\mathrm{S}$, $d_\mathrm{FM}$, $d_\mathrm{Ta}$ represent the reduced Planck's constant, spin-Hall angle, current in Ta layer, elementary charge, saturation magnetisation, sample width, thickness of NiFe and Ta layers,  respectively. 
$B_\mathrm{DL,1} = -B_\mathrm{DL,2}$ in symmetric stacks, hence the damping-like torques acting on $m$ and $n$ are listed:
\begin{align}
    \tau^\mathrm{DL}_m &= \gamma
    \begin{pmatrix}
        -B_\mathrm{DL}\,\langle n\rangle \sqrt{1-\langle n\rangle^2} \sin\phi \\
        0 \\
        B_\mathrm{DL}\,\langle n\rangle^2 \sin\phi
    \end{pmatrix}
    \label{eq:SI_tau_m_DL}
\end{align}
\begin{align}
    \tau^\mathrm{DL}_n &= \gamma
    \begin{pmatrix}
        -B_\mathrm{DL}\,\langle n\rangle \sqrt{1-\langle n\rangle^2} \cos\phi \\
        B_\mathrm{DL}\,(1-\langle n\rangle^2) \cos\phi \\
        0
        \label{eq:SI_tau_n_DL}
    \end{pmatrix}
\end{align}
According to Eqs.~\eqref{eq:SI_matrix_m}-\eqref{eq:SI_matrix_n} and Eqs.~\eqref{eq:SI_tau_m_FL}-\eqref{eq:SI_tau_n_DL}, the dynamic terms of acoustic mode $\delta m_{y'}$ and $\delta n_{x'}$ are derived into:
\begin{align}
    \delta m_{y'} &= 
    \frac{\gamma\big[\Omega^2 B_\mathrm{DL}\,\langle n\rangle\sqrt{1-\langle n\rangle^2}\sin\phi + i\Omega(\Omega_\mathrm{L}+\Omega_\mathrm{B}\sqrt{1-\langle n\rangle^2})(B_\mathrm{FL}\langle n\rangle\sin\phi)\big]}{-i\Omega(-\Omega_\mathrm{ac}+i\delta_\mathrm{ac}-\Omega)(\Omega_\mathrm{ac}+i\delta_\mathrm{ac}-\Omega)} \\
    \delta n_{x'} &= 
    \frac{\gamma\big[-\Omega^2 B_\mathrm{DL}\langle n\rangle^2\sin\phi - i\Omega B_\mathrm{FL}(\Omega_\mathrm{B}+2\Omega_\mathrm{ex})\langle n\rangle^2\sin\phi\big]}{-i\Omega(-\Omega_\mathrm{ac}+i\delta_\mathrm{ac}-\Omega)(\Omega_\mathrm{ac}+i\delta_\mathrm{ac}-\Omega)} \\
    \delta m_{x'} &= 
    \frac{\gamma\big[\Omega^2(B_\mathrm{DL}\langle n\rangle\sqrt{1-\langle n\rangle^2}\cos\phi) + i\Omega B_\mathrm{FL}\Omega_B \langle n\rangle\sqrt{1-\langle n\rangle^2}\cos\phi\big]}{-i\Omega(-\Omega_\mathrm{op}+i\delta_\mathrm{op}-\Omega)(\Omega_\mathrm{op}+i\delta_\mathrm{op}-\Omega)} \\
    \delta n_{y'} &= 
    \frac{\gamma\big[-\Omega^2 B_\mathrm{DL}(1-\langle n\rangle^2)\cos\phi - i\Omega B_\mathrm{FL}\Omega_B (1-\langle n\rangle^2)\cos\phi\big]}{-i\Omega(-\Omega_\mathrm{op}+i\delta_\mathrm{op}-\Omega)(\Omega_\mathrm{op}+i\delta_\mathrm{op}-\Omega)}
\end{align}
where $\tilde{\Omega}_\mathrm{ac(op)} = \pm \Omega_\mathrm{ac(op)} + i\delta_\mathrm{ac(op)}$. Calculating the residue at two poles ($- \Omega_\mathrm{ac(op)} + i\delta_\mathrm{ac(op)}$) and ($\Omega_\mathrm{ac(op)} + i\delta_\mathrm{ac(op)}$) using the residue theorem, $\mathrm{Res}(f,c) = \lim_{z\to c}(z-c)f(z)$ were obtained in simplified forms:
\begin{align}
    \delta m_{y'}(\Omega) &= \frac{\gamma}{\Omega_\mathrm{ac}+i\delta_\mathrm{ac}-\Omega}(a_{m_y'}+i b_{m_y'})
    + \frac{\gamma}{-\Omega_\mathrm{ac}+i\delta_\mathrm{ac}-\Omega}(-a_{my'}+i b_{my'}) \\
    \delta n_{x'}(\Omega) &= \frac{\gamma}{\Omega_\mathrm{ac}+i\delta_\mathrm{ac}-\Omega}(a_{n_x'}+i b_{n_x'})
    + \frac{\gamma}{-\Omega_\mathrm{ac}+i\delta_\mathrm{ac}-\Omega}(-a_{nx'}+i b_{nx'}) \\
    \delta m_{x'}(\Omega) &= \frac{\gamma}{\Omega_\mathrm{op}+i\delta_\mathrm{op}-\Omega}(a_{m_x'}+i b_{m_x'})
    + \frac{\gamma}{-\Omega_\mathrm{op}+i\delta_\mathrm{op}-\Omega}(-a_{m_x'}+i b_{m_x'}),\\
    \delta n_{y'}(\Omega) &= \frac{\gamma}{\Omega_\mathrm{op}+i\delta_\mathrm{op}-\Omega}(a_{n_y'}+i b_{n_y'})
    + \frac{\gamma}{-\Omega_\mathrm{op}+i\delta_\mathrm{op}-\Omega}(-a_{n_y'}+i b_{n_y'})
\end{align}
with the following coefficients:
\begin{align}
    a_{m_y'} &= \tfrac{1}{2} B_\mathrm{FL}\,\frac{\Omega_\mathrm{ac}}{\Omega_\mathrm{L}}\langle n\rangle\sin\phi \\
    b_{m_y'} &= -\tfrac{1}{2} B_\mathrm{DL}\langle n\rangle\sqrt{1-\langle n\rangle^2}\sin\phi\\
    a_{n_x'} &= -B_\mathrm{FL}\frac{(\Omega_\mathrm{B}+2\Omega_\mathrm{ex})}{2\Omega_\mathrm{ac}}\langle n\rangle^2\sin\phi\\
    b_{n_x'} &= \tfrac{1}{2} B_\mathrm{DL}\,\frac{\Omega_\mathrm{L}(\Omega_\mathrm{B}+2\Omega_\mathrm{ex})}{\Omega_\mathrm{ac}^2}\langle n\rangle^2\sqrt{1-\langle n\rangle^2}\sin\phi\\
    a_{m_x'} &= \tfrac{1}{2} B_\mathrm{FL}\,\frac{\Omega_B}{\Omega_\mathrm{op}}\langle n\rangle\sqrt{1-\langle n\rangle^2}\cos\phi \\
    b_{m_x'} &= -\tfrac{1}{2} B_\mathrm{DL}\langle n\rangle\sqrt{1-\langle n\rangle^2}\cos\phi\\
    a_{n_y'} &= -\tfrac{1}{2} B_\mathrm{FL}\,\frac{\Omega_B}{\Omega_\mathrm{op}}(1-\langle n\rangle^2)\cos\phi\\
    b_{n_y'} &= \tfrac{1}{2} B_\mathrm{DL}(1-\langle n\rangle^2)\cos\phi
\end{align}
Taking the magnetic resistance into account, the GMR effect is time-independent component due to $m_1$ and $m_2$ precess in phase for acoustic mode, while it is time-dependent for optical mode with out-of-phase precession. The time-varying resistances from MR effects are derived as following:
\begin{align}
    \Delta R_\mathrm{AMR}(t) &= \tfrac{1}{4}\Delta R_\mathrm{AMR}\Big[(m_1\cdot \hat{e}_{x})^2+(m_2\cdot \hat{e}_{x})^2\Big] \nonumber\\
    &= \tfrac{1}{2}\Delta R_\mathrm{AMR}\Big[\cos^2\phi(m_{x'}^2+n_{x'}^2)+\sin^2\phi(m_{y'}^2+n_{y'}^2)-2\sin\phi\cos\phi(m_{x'}m_{y'}+n_{x'}n_{y'})\Big] \nonumber\\
    &= \Delta R_\mathrm{AMR}\Big[\cos^2\phi \sqrt{1-\langle n\rangle^2}\delta m_{x'}(t) + \sin^2\phi \langle n\rangle\delta n_{y'}(t) -\sin\phi\cos\phi\Big (\sqrt{1-\langle n\rangle^2}\,\delta m_{y'}(t)+\langle n\rangle\delta n_{x'}(t)\Big)\Big]\\
    \Delta R_\mathrm{SMR}(t) &= \tfrac{1}{4}\Delta R_\mathrm{SMR}\Big[(m_1\cdot \hat{e}_{y})^2+(m_2\cdot \hat{e}_{y})^2\Big] \nonumber\\
    &= \tfrac{1}{2}\Delta R_\mathrm{SMR}\Big[\sin^2\phi(m_{x'}^2+n_{x'}^2)+\cos^2\phi(m_{y'}^2+n_{y'}^2)+2\sin\phi\cos\phi(m_{x'}m_{y'}+n_{x'}n_{y'})\Big] \nonumber\\
    &= \Delta R_\mathrm{SMR}\Big[\sin^2\phi \sqrt{1-\langle n\rangle^2}\,\delta m_{x'}(t) + \cos^2\phi \langle n\rangle\delta n_{y'}(t) +\sin\phi\cos\phi\Big( \sqrt{1-\langle n\rangle^2}\,\delta m_{y'}(t)+\langle n\rangle\delta n_{x'}(t)\Big) \Big]\\
    \Delta R_\text{GMR}(t) &= \tfrac{1}{2}\Delta R_\mathrm{GMR}(m_1\cdot m_2) \nonumber\\
    &= \tfrac{1}{2}\Delta R_\mathrm{GMR}\Big[(m_{x'}^2+m_{y'}^2)-(n_{x'}^2+n_{y'}^2)\Big] \nonumber\\
    &= \Delta R_\mathrm{GMR}\Big[ \sqrt{1-\langle n\rangle^2}\,\delta m_{x'}(t) - \langle n\rangle\delta n_{y'}(t) \Big]
\end{align}
The rf current $I(t)=I_\mathrm{rf}\cos\omega t$ along $x$ mixes with the resistance oscillation $\delta R(t)=\delta R\cos(\omega t+\varphi)$ given by Eqs.~(S29)--(S31), where $\varphi$ is the phase between resistance and current.
By Ohm's law, the time-averaged voltage is
\begin{equation}
V_\mathrm{dc}=\langle I(t)\,\delta R(t)\rangle=\tfrac{1}{2}\,I_\mathrm{rf}\,\delta R\cos\varphi .
\label{eq:Vdc}
\end{equation}
With $\delta R$ from Eqs.~(S17)--(S20), the $b$($a$) coefficients give (anti)symmetric Lorentzian peaks in $V_\mathrm{dc}$ as expressed below:
\begin{align}
    V^\mathrm{ac}_\mathrm{sym} &= -\frac{ \gamma}{2\delta_\mathrm{ac}}R_0(\Delta R_\mathrm{AMR}-\Delta R_\mathrm{SMR})\,I^2_\mathrm{rf}\frac{d}{dI}(b_{m_y'}\sqrt{1-\langle n\rangle^2}\sin\phi\cos\phi+b_{n_x'}\langle n\rangle\sin\phi\cos\phi) \nonumber\\
    &= \frac{I_\mathrm{rf} R_0 B_\mathrm{DL}}{8\Delta B_\mathrm{ac}\sqrt{1+\frac{B_\mathrm S}{2B_\mathrm{ex}}}}\, (\Delta R_\mathrm{AMR}-\Delta R_\mathrm{SMR})\,\langle n\rangle (1-2\langle n\rangle^2) \sin 2\phi \sin\phi  \label{eq:SI_Vsym_ac_all}\\
    V^\mathrm{ac}_\mathrm{asy} &= -\frac{ \gamma}{2\delta_\mathrm{ac}}R_0(\Delta R_\mathrm{AMR}-\Delta R_\mathrm{SMR})\,I^2_\mathrm{rf}\frac{d}{dI}(a_{m_y'}\sqrt{1-\langle n\rangle^2}\sin\phi\cos\phi+a_{n_x'}\langle n\rangle\sin\phi\cos\phi) \nonumber\\
    &= -\frac{I_\mathrm{rf} R_0 B_\mathrm{FL}}{8\Delta B_\mathrm{ac}}(\Delta R_\mathrm{AMR}-\Delta R_\mathrm{SMR})\,\frac{\langle n\rangle}{\sqrt{1-\langle n\rangle^2} }(1-2\langle n\rangle^2) \sin 2\phi \sin\phi  \label{eq:SI_Vasy_ac_all}\\
    V^\mathrm{op}_\mathrm{sym} &= \frac{\gamma}{2\delta_\mathrm{op}}R_0\,I^2_\mathrm{rf}\frac{d}{dI}\Big\{\Big[\tfrac{1}{2}(\Delta R_\mathrm{AMR}-\Delta R_\mathrm{SMR})\cos 2\phi+\Delta R_\mathrm{GMR}\Big]\big(b_{m_x'}\sqrt{1-\langle n\rangle^2}-b_{n_y'}\langle n\rangle\big)\Big\} \nonumber\\
    &= -\frac{I_\mathrm{rf} R_0 B_\mathrm{DL}}{4\Delta B_\mathrm{op}} \sqrt{\frac{2B_\mathrm{ex}}{B_\mathrm{S}}}\, \langle n\rangle |\langle n\rangle| \sqrt{1-\langle n\rangle^2}\big[(\Delta R_\mathrm{AMR}-\Delta R_\mathrm{SMR})\cos2\phi\cos\phi\,+2\Delta R_\mathrm{GMR}\cos\phi \big]  \label{eq:SI_Vsym_op_all}\\  
    V^\mathrm{op}_\mathrm{asy} &= \frac{\gamma}{2\delta_\mathrm{op}}R_0\,I^2_\mathrm{rf}\frac{d}{dI}\Big\{\Big[\tfrac{1}{2}(\Delta R_\mathrm{AMR}-\Delta R_\mathrm{SMR})\cos 2\phi+\Delta R_\mathrm{GMR}\Big]\big(a_{m_x'}\sqrt{1-\langle n\rangle^2}-a_{n_y'}\langle n\rangle\big)\Big\} \nonumber\\
    &= -\frac{I_\mathrm{rf} R_0 B_\mathrm{FL}}{4\Delta B_\mathrm{op}}\,\langle n\rangle\sqrt{1-\langle n\rangle^2} \big[(\Delta R_\mathrm{AMR}-\Delta R_\mathrm{SMR}) \cos 2\phi \cos\phi + 2\Delta R_\mathrm{GMR}\cos\phi\big]  \label{eq:SI_Vasy_op_all}
\end{align}
where $I_\mathrm{rf}$, $R_0$, $\Delta B_\mathrm{ac(op)}$ are the microwave current flowing through the sample, static resistance, and the half width at half maximum (HWHM) linewidth of each branch, respectively. The frequency linewidth $\delta_\mathrm{ac(op)}$ is converted to field-swept linewidth $\Delta B_\mathrm{ac(op)}$ from the relation $\Delta B_\mathrm{ac(op)}=\frac{\delta_\mathrm{ac(op)}}{|d\Omega_\mathrm{ac(op)}/dB|}$, where $d\Omega_\mathrm{ac}/dB = \gamma \sqrt{1+B_\mathrm{S}/(2B_\mathrm{ex})}$, $d\Omega_\mathrm{op}/dB = -\gamma \sqrt{2B_\mathrm{S}B_\mathrm{ex}}\cdot\frac{B}{4B^2_\mathrm{ex}|\langle n\rangle|}$.

\newpage

\section*{Supplementary Note 2: Power dependence of magnetoresistance effects}

\subsection{Calibration of microwave power}
\begin{figure}[ht!]
    \centering
    \includegraphics[width=0.75\linewidth]{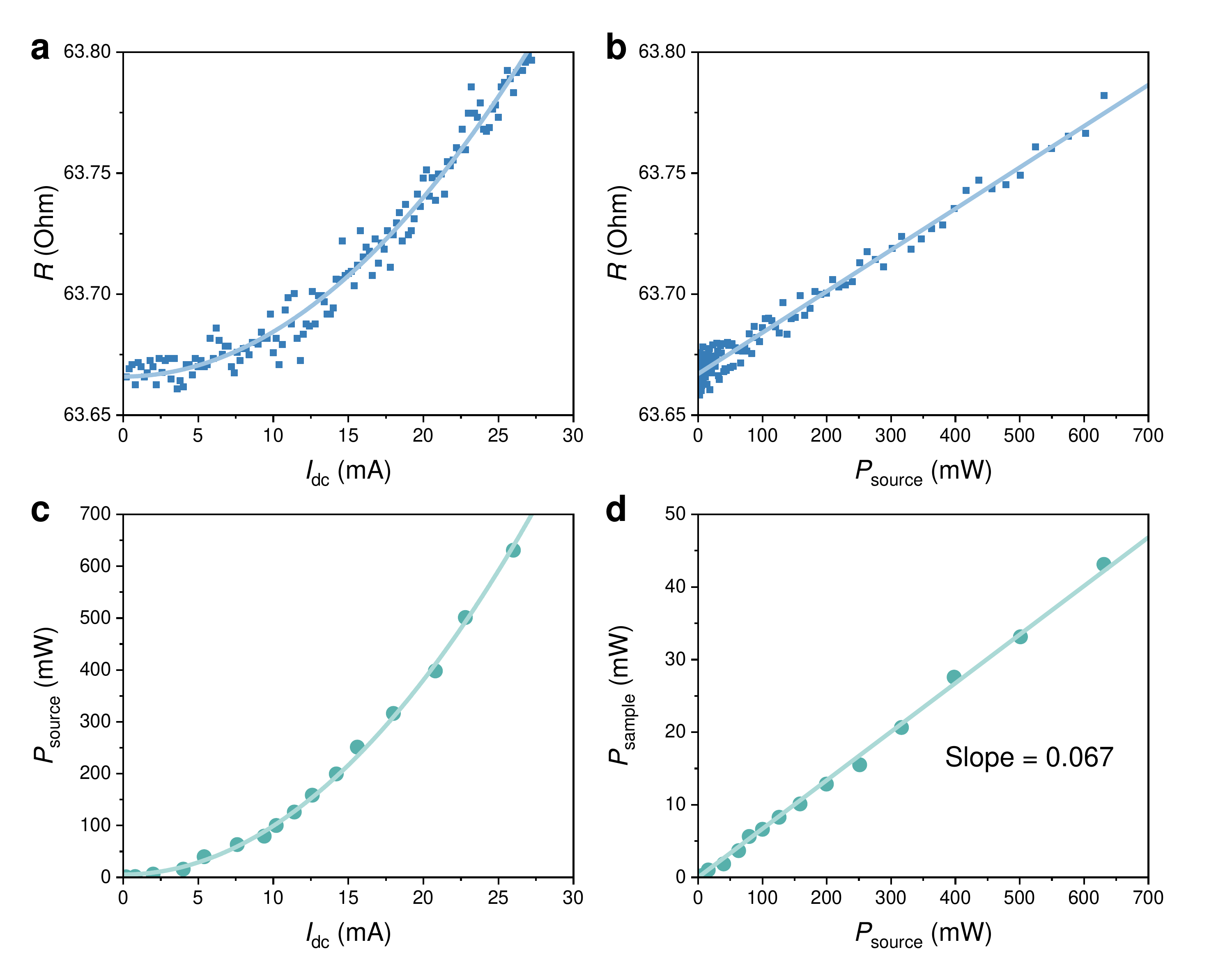}
    \caption{\textbf{Microwave calibration.} Resistance change as a function of \textbf{a}, applied dc current, \textbf{b}, output power of microwave source. \textbf{c}, Current calibration of power from microwave source. \textbf{d}, Power in sample $P_\mathrm{sample}$ as a function of power from source $P_\mathrm{source}$.}
    \label{fig:SI_mw_calibration}
\end{figure}

Only a small fraction of microwave power could be transported from microwave source to the device due to the inherent impedance mismatch and circuit loss in practical experimental system, therefore a bolometric technique\cite{Fang2011,sud2021parity} is adopted to quantify the actual power $P_\mathrm{sample}$ flowing through the device. The Joule heating induced resistance change between the output power of microwave source $P_\mathrm{source}$ (Fig.~\ref{fig:SI_mw_calibration}b) and a certain applied dc current $I_\mathrm{dc}$ (Fig.~\ref{fig:SI_mw_calibration}a) is compared, the current calibration of $P_\mathrm{source}$ is presented in Fig.~\ref{fig:SI_mw_calibration}(c). A power transport ratio $P_\mathrm{sample}/P_\mathrm{source}$ of 6.7\% is extracted from Fig.~\ref{fig:SI_mw_calibration}(d).

\subsection{Power dependence of FMR voltage}
The linear relation between FMR voltage and microwave power can be confirmed by the rewritten Eqs.~\eqref{eq:SI_Vsym_ac_all}-\eqref{eq:SI_Vasy_op_all}:
\begin{align}
    V_\mathrm{sym(asy)}^\mathrm{ac}(\phi) &= A_\mathrm{sym(asy)}^\mathrm{ac}(\phi)\cdot(\Delta R_\mathrm{AMR}-\Delta R_\mathrm{SMR})\sin 2\phi \sin\phi\cdot P_\mathrm{mw}  \\
    V_\mathrm{sym(asy)}^\mathrm{op}(\phi) &= A_\mathrm{sym(asy)}^\mathrm{op}(\phi)\cdot\big((\Delta R_\mathrm{AMR}-\Delta R_\mathrm{SMR})\cos2\phi\cos\phi + 2\Delta R_\mathrm{GMR}\cos\phi\big)\cdot P_\mathrm{mw}
\end{align}
where $A_\mathrm{sym}^\mathrm{ac}(\phi)=\frac{B_\mathrm{DL}\,}{8 I_\mathrm{rf}\Delta B_\mathrm{ac}\sqrt{1+\frac{B_\mathrm S}{2B_\mathrm{ex}}}}\langle n\rangle (1-2\langle n\rangle^2)$,
$A_\mathrm{asy}^\mathrm{ac}(\phi)=-\frac{B_\mathrm{FL}\,}{8 I_\mathrm{rf}\Delta B_\mathrm{ac}}\frac{\langle n\rangle (1-2\langle n\rangle^2)}{\sqrt{1-\langle n\rangle^2}}$, 
$A_\mathrm{sym}^\mathrm{op}(\phi)=-\frac{B_\mathrm{DL}}{4I_\mathrm{rf}\Delta B_\mathrm{op}} \sqrt{\frac{2B_\mathrm{ex}}{B_\mathrm{S}}}\, \langle n\rangle |\langle n\rangle| \sqrt{1-\langle n\rangle^2}$, 
$A_\mathrm{asy}^\mathrm{op}(\phi)=-\frac{B_\mathrm{FL}}{4I_\mathrm{rf}\Delta B_\mathrm{op}}\,\langle n\rangle\sqrt{1-\langle n\rangle^2}$, 
$P_\mathrm{mw}$ is the actual power flowing through the sample $P_\mathrm{sample}$.

In addition, Eqs.~\eqref{eq:SI_Vsym_ac_all}-\eqref{eq:SI_Vasy_op_all} can be rearranged into another expressions as following:
\begin{align}
    V_\mathrm{sym(asy)}^\mathrm{ac}(\phi,P_\mathrm{mw}) &= V_\mathrm{A-S}^\mathrm{ac,sym(asy)}(P_\mathrm{mw}) \sin 2\phi \sin\phi   \\
    \frac{V_\mathrm{sym(asy)}^\mathrm{op}(\phi,P_\mathrm{mw})}{\cos\phi} &= V_\mathrm{A-S}^\mathrm{op,sym(asy)}(P_\mathrm{mw})\cos2\phi + V_\mathrm{G}^\mathrm{sym(asy)}(P_\mathrm{mw})    \label{eq:SI_V_op_phi}
\end{align}
where $V_\mathrm{A-S}^\mathrm{ac,sym}=\frac{I_\mathrm{rf} R_0 B_\mathrm{DL}}{8\Delta B_\mathrm{ac}\sqrt{1+\frac{B_\mathrm S}{2B_\mathrm{ex}}}}\,\langle n\rangle (1-2\langle n\rangle^2)\cdot (\Delta R_\mathrm{AMR}-\Delta R_\mathrm{SMR})$,
$V_\mathrm{A-S}^\mathrm{ac,asy}=-\frac{I_\mathrm{rf} R_0 B_\mathrm{FL}\,}{8\Delta B_\mathrm{ac}}\frac{\langle n\rangle (1-2\langle n\rangle^2)}{\sqrt{1-\langle n\rangle^2}}\cdot(\Delta R_\mathrm{AMR}-\Delta R_\mathrm{SMR})$, 
$V_\mathrm{A-S}^\mathrm{op,sym}= -\frac{I_\mathrm{rf} R_0 B_\mathrm{DL}}{4\Delta B_\mathrm{op}} \sqrt{\frac{2B_\mathrm{ex}}{B_\mathrm{S}}}\, \langle n\rangle |\langle n\rangle| \sqrt{1-\langle n\rangle^2}\cdot(\Delta R_\mathrm{AMR}-\Delta R_\mathrm{SMR})$, 
$V_\mathrm{G}^\mathrm{sym}=-\frac{I_\mathrm{rf} R_0 B_\mathrm{DL}}{4\Delta B_\mathrm{op}} \sqrt{\frac{2B_\mathrm{ex}}{B_\mathrm{S}}}\,\langle n\rangle |\langle n\rangle| \sqrt{1-\langle n\rangle^2}\cdot2\Delta R_\mathrm{GMR}$, 
$V_\mathrm{A-S}^\mathrm{op,asy}= -\frac{I_\mathrm{rf} R_0 B_\mathrm{FL}}{4\Delta B_\mathrm{op}}\, \langle n\rangle \sqrt{1-\langle n\rangle^2}\cdot(\Delta R_\mathrm{AMR}-\Delta R_\mathrm{SMR})$,
and $V_\mathrm{G}^\mathrm{asy}=-\frac{I_\mathrm{rf} R_0 B_\mathrm{FL}}{4\Delta B_\mathrm{op}}\,\langle n\rangle\sqrt{1-\langle n\rangle^2}\cdot2\Delta R_\mathrm{GMR}$.
Based on the expressions above, the FMR voltage associated with AMR and SMR effects $V_\mathrm{A-S}$ or GMR effect $V_\mathrm{G}$ can be conducted by fitting the power dependence of FMR spectra of either symmetric or antisymmetric component as a function of applied magnetic field angle $\phi$, or solving the equations of  FMR amplitude at specific $\phi$ angles.

\section*{Supplementary Note 3: GMR rectification of current-induced N\'eel vector reversal}
\begin{figure}[hb!]
    \centering
    \includegraphics[width=1.0\linewidth]{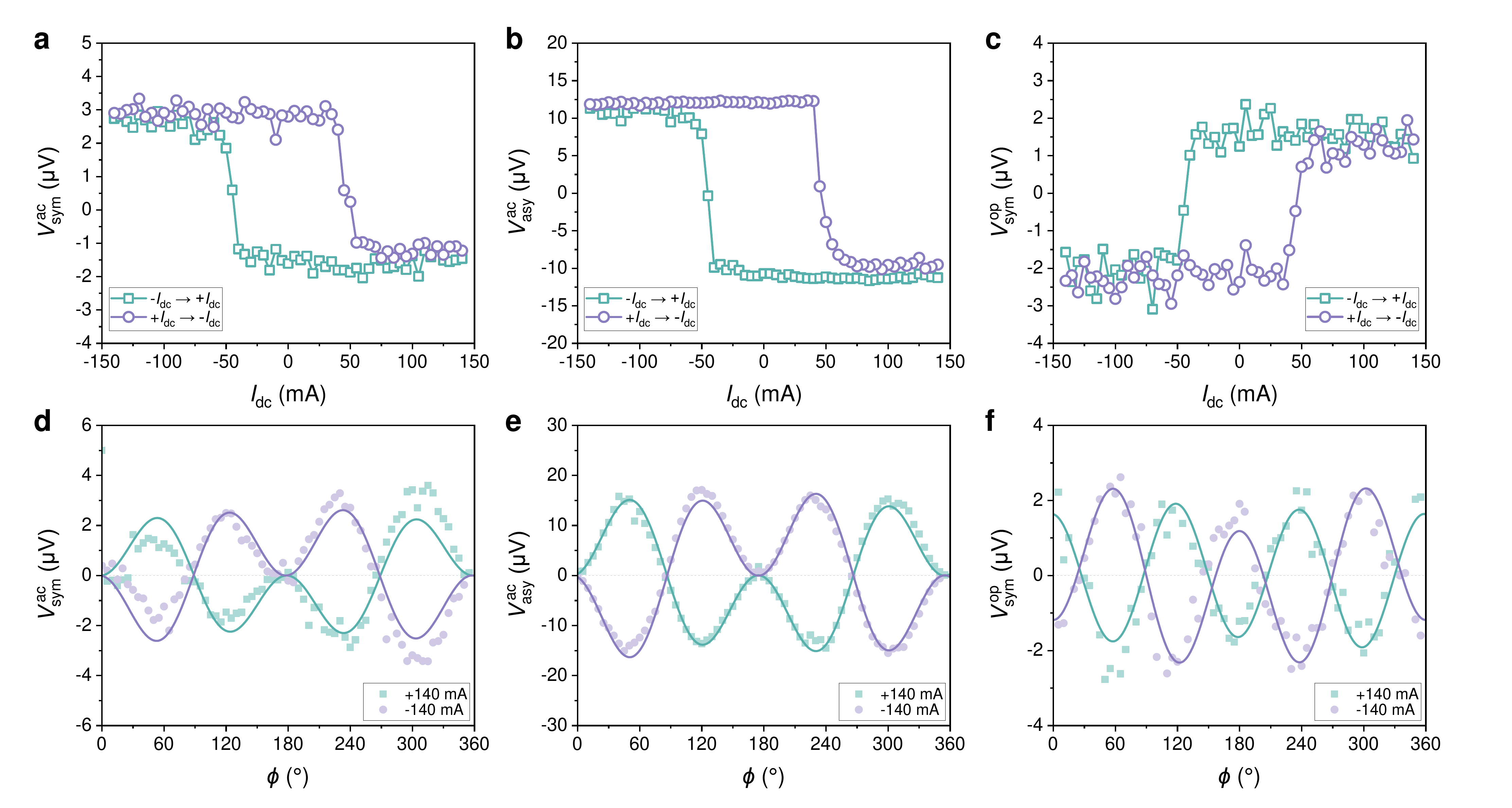}
    \caption{\textbf{ST-FMR spectra of current-induced N\'eel vector reversal.} The amplitude of symmetric $V_\mathrm{sym}^\mathrm{ac}$ (\textbf{a}) and antisymmetric $V_\mathrm{asy}^\mathrm{ac}$ (\textbf{b}) components for acoustic mode, and symmetric component for optical mode $V_\mathrm{sym}^\mathrm{op}$ (\textbf{c}) for each FMR voltage trace, respectively. Angular dependence of $V_\mathrm{sym}^\mathrm{ac}$ (\textbf{d}), $V_\mathrm{asy}^\mathrm{ac}$ (\textbf{e}), $V_\mathrm{sym}^\mathrm{op}$ (\textbf{f}) as a function of applied magnetic field angle $\phi$ at dc current $I_\mathrm{dc}$ of $\pm140$\,mA.}
    \label{fig:SI_switch_sym}
\end{figure}

$V_\mathrm{A-S}$ and $V_\mathrm{G}$ can be calculated from $V_\mathrm{sym}^\mathrm{op}$ for two angles such as $\phi=120^\circ$ and $140^\circ$ using Eq.~\eqref{eq:SI_V_op_phi}:

\begin{align}
    \frac{V_\mathrm{sym}^\mathrm{op}(120^\circ,P_\mathrm{mw})}{-0.5} &= -0.5\,V_\mathrm{A-S}(120^\circ,P_\mathrm{mw}) + V_\mathrm{G}(120^\circ,P_\mathrm{mw})\\
    \frac{V_\mathrm{sym}^\mathrm{op}(140^\circ,P_\mathrm{mw})}{-0.766} &= 0.174\,V_\mathrm{A-S}(140^\circ,P_\mathrm{mw}) + V_\mathrm{G}(140^\circ,P_\mathrm{mw})
\end{align}
or for antisymmetric component:
\begin{align}
    \frac{V_\mathrm{asy}^\mathrm{op}(120^\circ,P_\mathrm{mw})}{-0.5} &= -0.5\,V_\mathrm{A-S}(120^\circ,P_\mathrm{mw}) + V_\mathrm{G}(120^\circ,P_\mathrm{mw})\\
    \frac{V_\mathrm{asy}^\mathrm{op}(140^\circ,P_\mathrm{mw})}{-0.766} &= 0.174\,V_\mathrm{A-S}(140^\circ,P_\mathrm{mw}) + V_\mathrm{G}(140^\circ,P_\mathrm{mw})
\end{align}

\begin{figure}[ht!]
    \centering
    \includegraphics[width=1.0\linewidth]{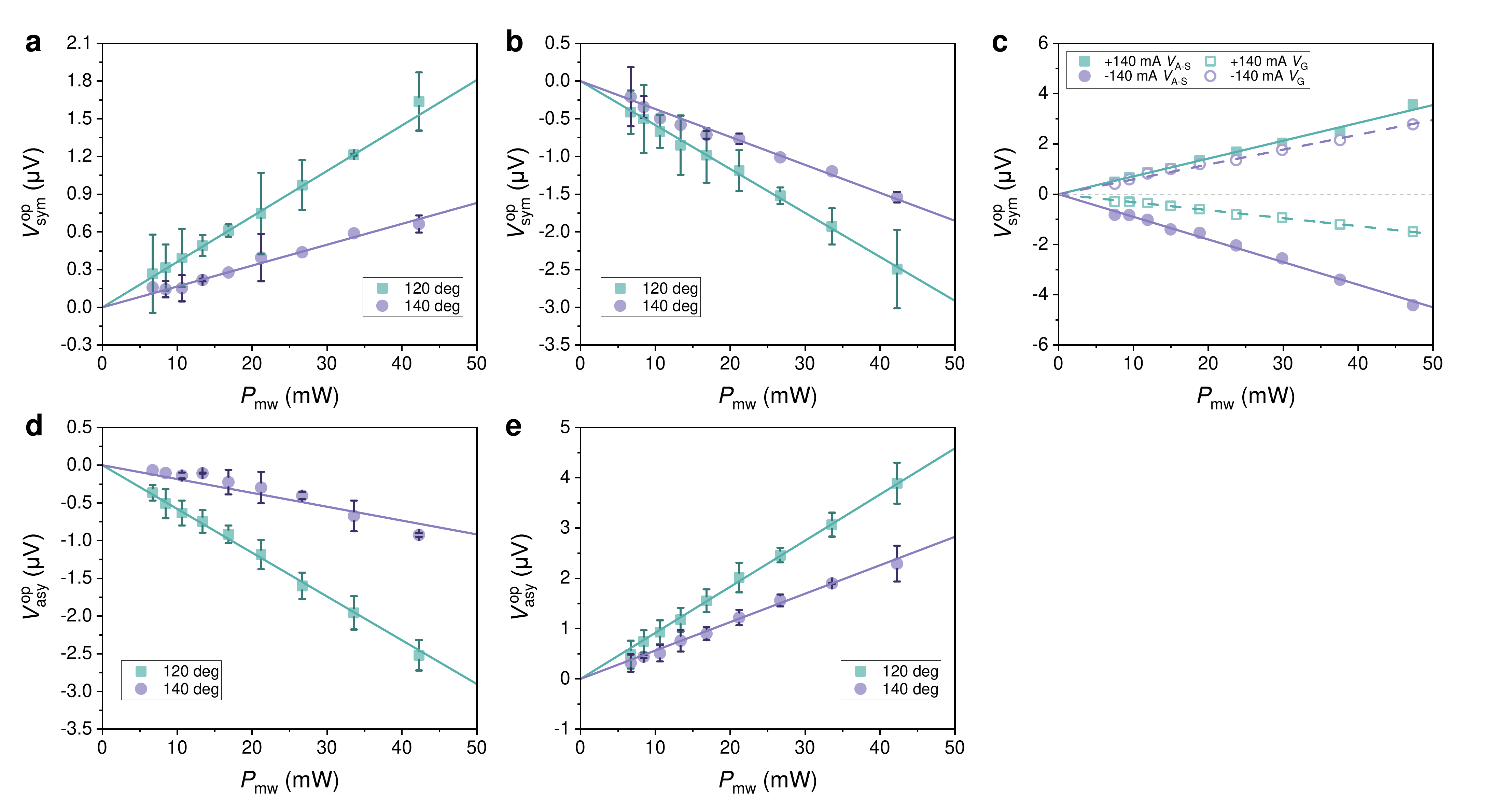}
    \caption{\textbf{GMR rectification of current-induced N\'eel vector reversal.} Microwave power dependence of symmetric component of FMR voltage $V_\mathrm{sym}^\mathrm{op}$ for $I_\mathrm{dc}=+140\,\mathrm{mA}$ (\textbf{a}) and $I_\mathrm{dc}=-140\,\mathrm{mA}$ (\textbf{b}), and antisymmetric component of FMR voltage $V_\mathrm{asy}^\mathrm{op}$ for $I_\mathrm{dc}=+140\,\mathrm{mA}$ (\textbf{d}) and $I_\mathrm{dc}=-140\,\mathrm{mA}$ (\textbf{e}) at field angle $\phi=120^\circ$ and $140^\circ$. \textbf{c}, Applied dc current $I_\mathrm{dc}$ dependence of $V_\mathrm{sym}^\mathrm{op}$ as a function of microwave power $P_\mathrm{mw}$ fitting by separated MR effects.}
    \label{fig:SI_dc_mw_sym}
\end{figure}

\newpage
\clearpage

\section*{Supplementary Note 4: Other supplementary figures}
\subsection{FMR voltage rectifications as a function of power}
\begin{figure}[ht!]
    \centering
    \includegraphics[width=0.75\linewidth]{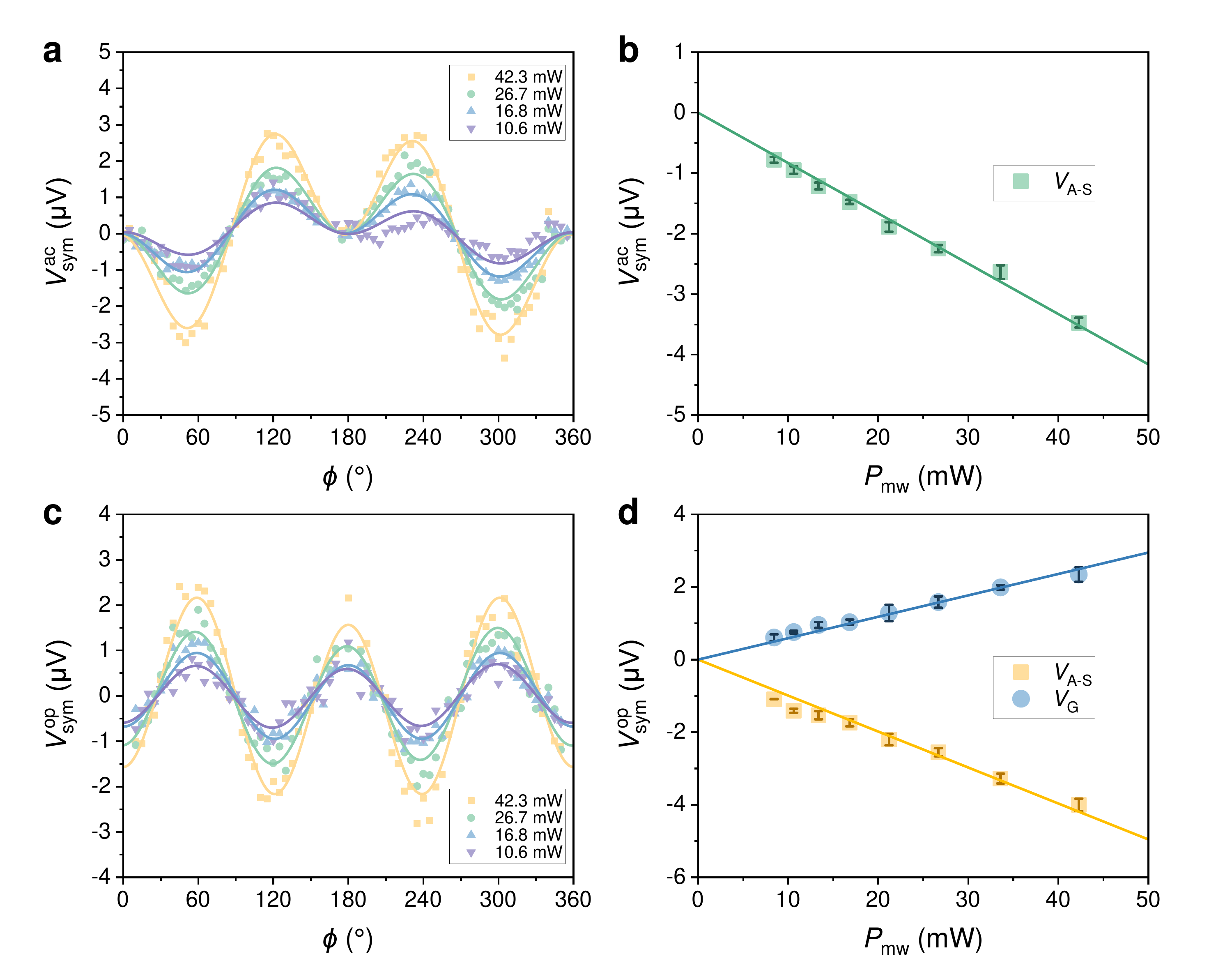}
    \caption{Microwave power dependent symmetric component of FMR voltage $V_\mathrm{sym}$ as a function of field angle $\phi$ for acoustic (\textbf{a}) and optical (\textbf{c}) modes. Separated MR effect dependence of $V_\mathrm{sym}$ as a function of microwave power $P_\mathrm{mw}$ for acoustic (\textbf{b}) and optical (\textbf{d}) modes.}
    \label{fig:SI_mw_sym}
\end{figure}
\newpage
\subsection{GMR rectification via in-plane magnetic resonance}
\begin{figure}[ht!]
    \centering
    \includegraphics[width=0.75\linewidth]{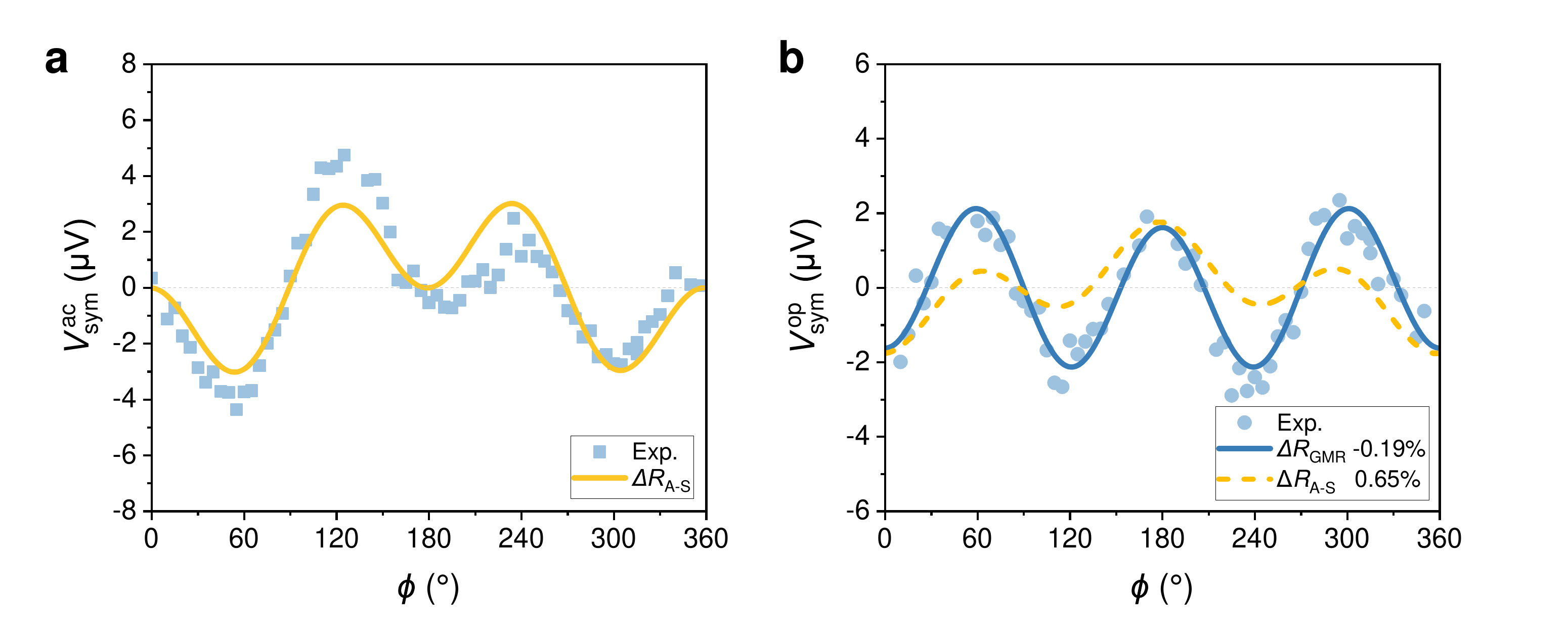}
    \caption{\textbf{GMR rectification via in-plane magnetic resonance.} Symmetric component of FMR voltage $V_\mathrm{sym}$ as a function of magnetic field angle $\phi$ for acoustic (\textbf{a}) and optical (\textbf{b}) modes fitting without/with GMR component.}
    \label{fig:SI_FMR_sym}
\end{figure}
\begin{figure}[ht!]
    \centering
    \includegraphics[width=0.5\linewidth]{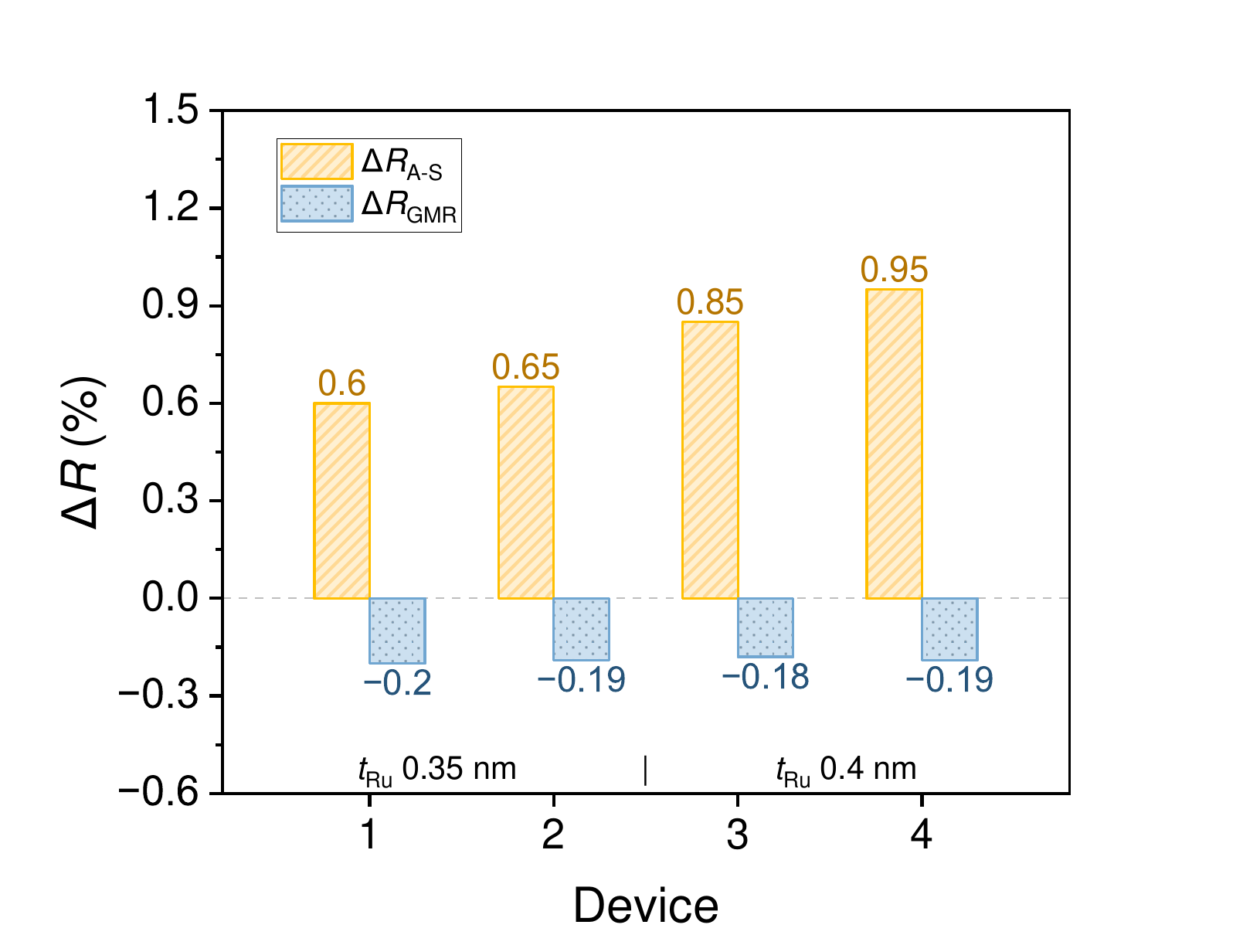}
    \caption{\textbf{Comparison of Ru thickness.} The MR ratios calculated from angular dependence of FMR voltage $V_\mathrm{asy}^\mathrm{op}$ for varying spacer layer (Ru) thickness, device 1 and 2 for 0.35\,nm, device 3 and 4 for 0.4\,nm, respectively. Here, device 2 is the one discussed in this work.}
    \label{fig:SI_device_vs}
\end{figure}
\newpage
\subsection{Experimental data and calculated results of magnetoresistance}
\begin{figure}[ht!]
    \centering
    \includegraphics[width=0.75\linewidth]{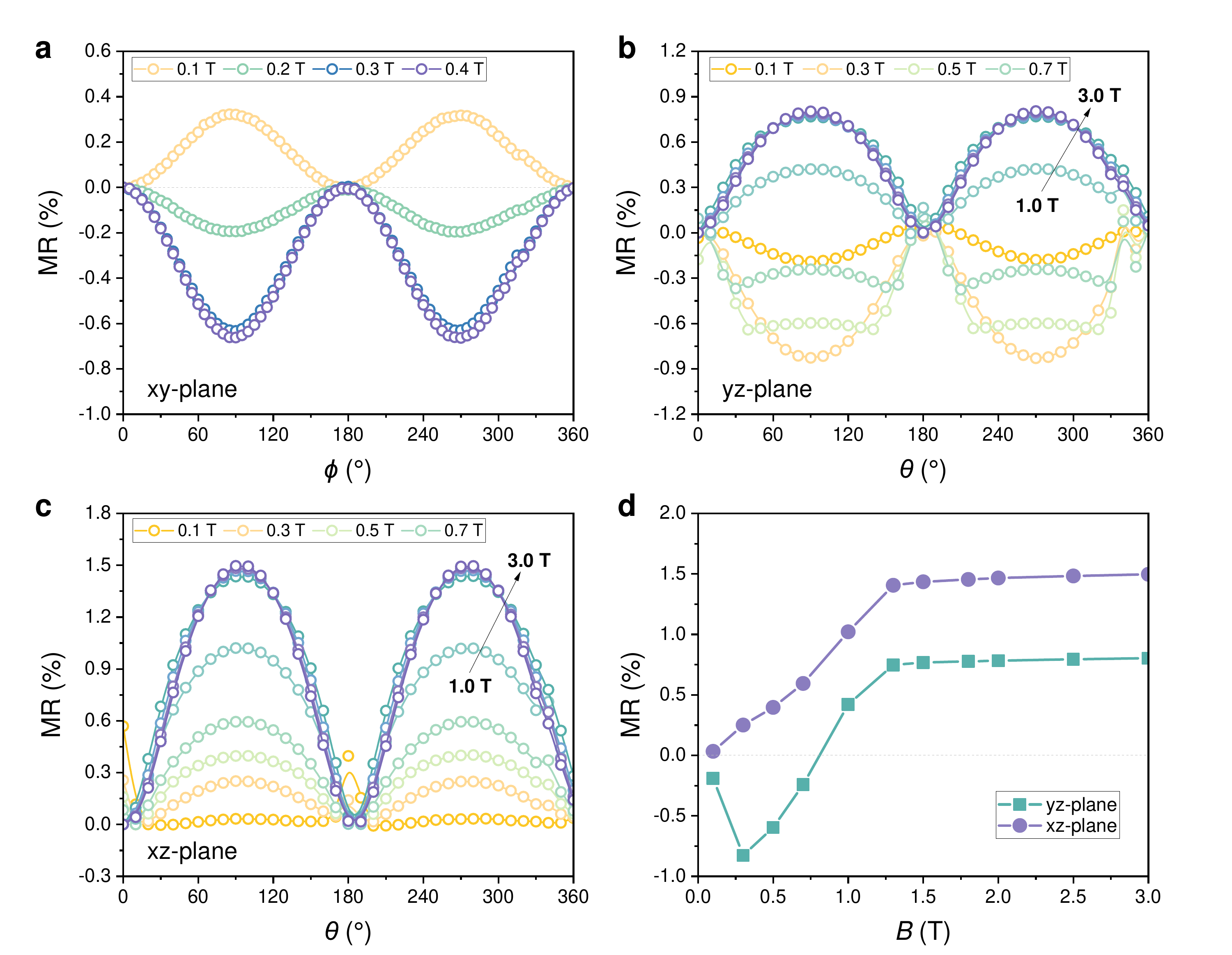}
    \caption{\textbf{Angular dependence of MR curves.} Magnetic field dependence of magnetoresistance as a function of \textbf{a}, in-plane field angle $\phi$, and out-of-plane field angle $\theta$ with transverse (\textbf{b}) and longitudinal (\textbf{c}) rotation. \textbf{d}, Out-of-plane magnetic field dependence of MR ratio.}
    \label{fig:SI_OOP_MR}
\end{figure}

\begin{figure}[ht!]
    \centering
    \includegraphics[width=0.75\linewidth]{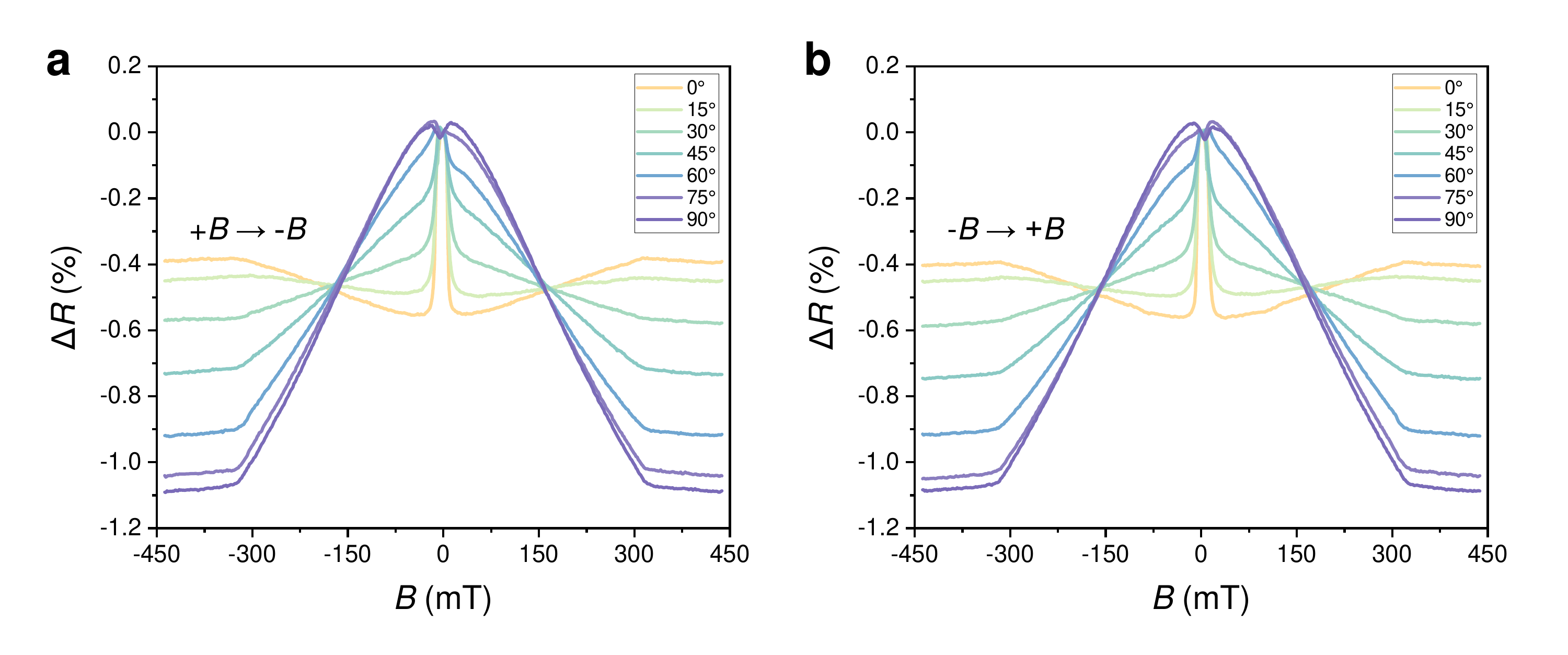}
    \caption{\textbf{Field dependence of MR curves.} Angular dependence of magnetoresistance as a function of forward (\textbf{a}) and backward (\textbf{b}) sweep of applied magnetic field. }
    \label{fig:SI_field_MR}
\end{figure}
\newpage
\begin{figure}[ht!]
    \centering
    \includegraphics[width=1.0\linewidth]{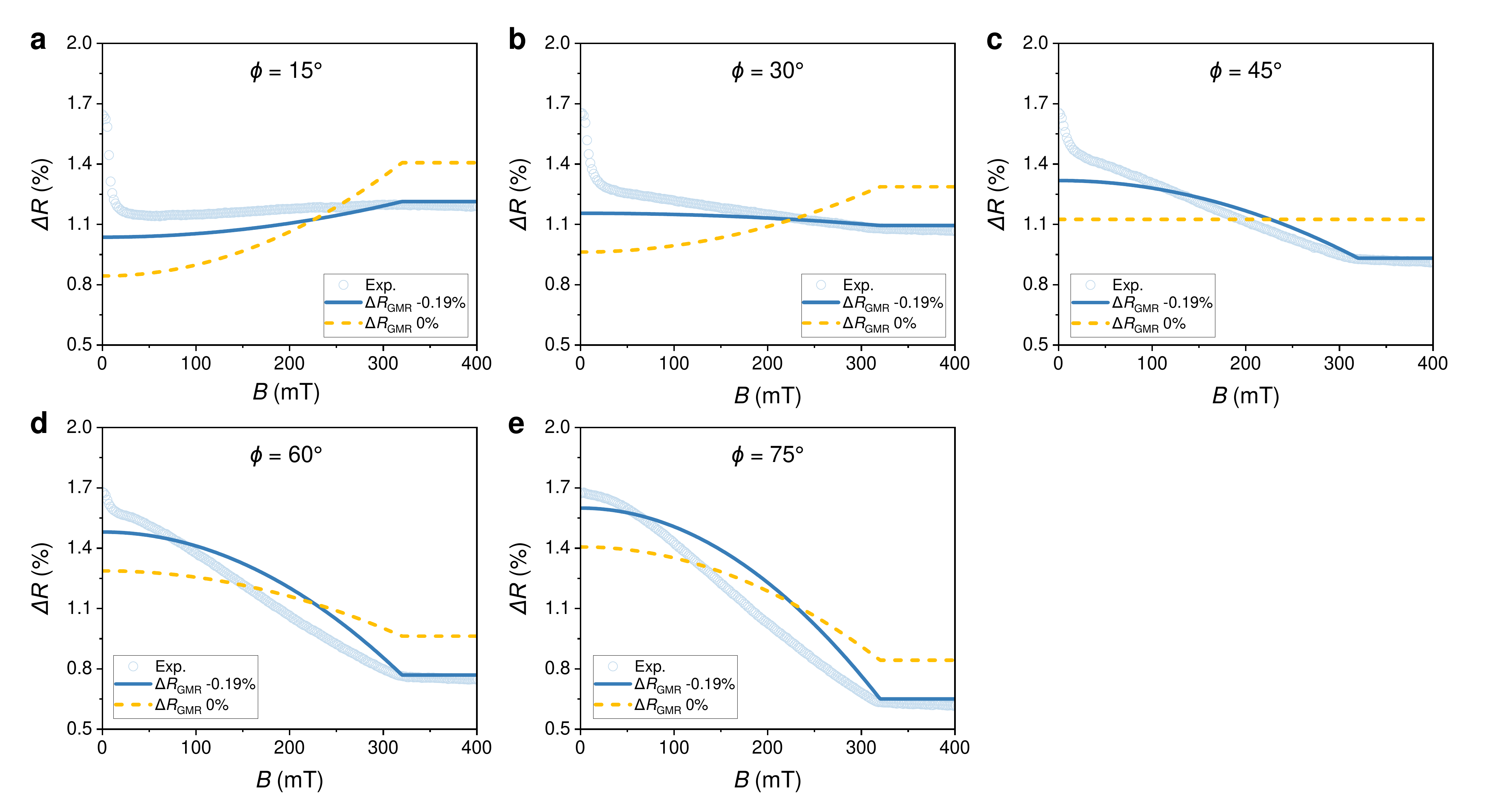}
    \caption{\textbf{GMR contribution to the in-plane magnetoresistance.}
    Measured MR (symbols) and macrospin calculations with $\Delta R_\mathrm{GMR}=-0.19\%$ (solid lines) and $0$ (dashed lines) at field angles $\phi$ of \textbf{a}, $15^\circ$, \textbf{b}, $30^\circ$, \textbf{c}, $45^\circ$, \textbf{d}, $60^\circ$ and \textbf{e}, $75^\circ$.
The calculation places the two moments at $\phi\pm\phi_\mathrm{c}$, with $\cos\phi_\mathrm{c}=B/2B_\mathrm{ex}$ below saturation, so that
$\Delta R=\tfrac{1}{2}(\Delta R_\mathrm{AMR}+\Delta R_\mathrm{SMR})+\tfrac{1}{2}(\Delta R_\mathrm{AMR}-\Delta R_\mathrm{SMR})\cos2\phi\cos2\phi_\mathrm{c}+\Delta R_\mathrm{GMR}\cos2\phi_\mathrm{c}$.
At $\phi=45^\circ$ the AMR and SMR terms are independent of field, because one layer rotates towards the current and the other away from it by the same angle, so the field dependence arises from GMR alone.
The model neglects the weak shape anisotropy along $x$, which aligns the moments with $x$ near zero field and is the likely origin of the deviations at low field.}
    \label{fig:SI_MR_calc}
\end{figure}

\end{document}